\documentclass[reprint,aps, pra,superscriptaddress, floatfix]{revtex4-2}
\usepackage{amsmath}
\usepackage{bm}
\usepackage{graphicx}
\usepackage[caption=false]{subfig}
\usepackage{lmodern}
\usepackage{amsmath}
\usepackage{physics}
\usepackage{natbib}
\usepackage{tikz}
\usetikzlibrary{arrows.meta}
\usepackage{color}
\usepackage{diagbox}
\usepackage{bm}
\usepackage{amssymb}
\usepackage{hyperref}
\usepackage{xcolor}
\hypersetup{
    pdftoolbar=true,        
    pdfmenubar=true,        
    pdffitwindow=false,     
    pdfstartview={FitH},    
    pdftitle={My title},    
    pdfauthor={Author},     
    pdfsubject={Subject},   
    pdfcreator={Creator},   
    pdfproducer={Producer}, 
    pdfkeywords={keyword1} {key2} {key3}, 
    pdfnewwindow=true,      
    colorlinks=true,       
    linkcolor=carmine,          
    citecolor=carmine,        
}

\renewcommand{\(}{\left(}
\renewcommand{\)}{\right)}
\renewcommand{\[}{\left[}
\renewcommand{\]}{\right]}

\newcommand{\mc}{\mathcal}

\definecolor{vermillion}{rgb}{0.86, 0.18, 0.01}
\definecolor{AJ}{rgb}{0.0, 0.48, 0.65}
\definecolor{carmine}{rgb}{0.59, 0.0, 0.09}
\definecolor{d_blue}{cmyk}{0.91, 0.79, 0.00, 0.22}
\definecolor{go_green}{rgb}{0.13, 0.55, 0.13}
\definecolor{sohaib}{rgb}{0.90, 0.10, 0.10}
\definecolor{andy}{rgb}{0.60, 0.10, 0.10}

\newcommand{\reportnumber}{FERMILAB-PUB-22-083-SQMS}
\usepackage[angle=0,scale=1,color=black,firstpage=true,opacity=1]{background}

\SetBgContents{\small\texttt{\reportnumber}}
\SetBgPosition{current page.north east}
\SetBgHshift{-3.5cm}
\SetBgVshift{-1.5cm}

\begin{document}
\title{Fermionic approach to variational quantum simulation of Kitaev spin models}

\author{Ammar Jahin}
\affiliation{Department of Physics, University of Florida, 2001 Museum Rd, Gainesville, FL 32611, USA}
\author{Andy C.~Y.~Li}
\affiliation{Fermi National Accelerator Laboratory, Batavia, IL, 60510, USA}
\author{Thomas Iadecola}
\affiliation{Department of Physics and Astronomy, Iowa State University, Ames, Iowa 50011, USA}
\affiliation{Ames Laboratory, Ames, Iowa 50011, USA}
\author{Peter P. Orth}
\affiliation{Department of Physics and Astronomy, Iowa State University, Ames, Iowa 50011, USA}
\affiliation{Ames Laboratory, Ames, Iowa 50011, USA}
\author{Gabriel N. Perdue}
\affiliation{Fermi National Accelerator Laboratory, Batavia, IL, 60510, USA}
\author{Alexandru Macridin}
\affiliation{Fermi National Accelerator Laboratory, Batavia, IL, 60510, USA}
\author{M. Sohaib Alam}
\affiliation{Quantum Artificial Intelligence Lab. (QuAIL), Exploration Technology Directorate, NASA Ames Research Center, Moffett Field, CA 94035, USA}
 \affiliation{USRA Research Institute for Advanced Computer Science (RIACS), Mountain View, CA, 94043, USA}
\author{Norm M. Tubman}
\affiliation{Quantum Artificial Intelligence Lab. (QuAIL), Exploration Technology Directorate, NASA Ames Research Center, Moffett Field, CA 94035, USA}

\date{\today}
\begin{abstract}
    We use the variational quantum eigensolver (VQE) to simulate Kitaev spin models with and without integrability breaking perturbations, focusing in particular on the honeycomb and square-octagon lattices. 
    These models are well known for being exactly solvable in a certain parameter regime via a mapping to free fermions.
    We use classical simulations to explore a novel variational ansatz that takes advantage of this fermionic representation and is capable of expressing the exact ground state in the solvable limit. 
    We also demonstrate that this ansatz can be extended beyond this limit to provide excellent accuracy when compared to other VQE approaches. 
    In certain cases, this fermionic representation is advantageous because it reduces by a factor of two the number of qubits required to perform the simulation. 
    We also comment on the implications of our results for simulating non-Abelian anyons on quantum computers.
\end{abstract}

\maketitle

\tableofcontents

\section{Introduction}
One of the hallmarks of frustrated interacting two-dimensional quantum spin systems is the emergence of quantum spin liquid ground states with long-range topological order and fractionalized excitations that obey (non-)Abelian statistics~\cite{balentsSpinLiquidsFrustrated2010,savaryQuantumSpinLiquids2017}. 
The celebrated Kitaev spin model~\cite{Kitaev_2006}, which describes spins on a trivalent lattice interacting via an anisotropic Ising coupling, is a popular playground for theoretically studying such phenomena. 
The model is exactly solvable in terms of fermionic operators, meaning the Hamiltonian reduces to a quadratic form.
This means that many properties of the model can be analytically obtained either exactly or within the framework of perturbation theory~\cite{Kitaev_2006, Yang_Zhou_Sun_2007, baskaranExactResultsSpin2007,hermannsPhysicsKitaevModel2018}.
From a computational perspective being quadratic means that for a system of $N$ spins, one only needs to diagonalize an $N \times N$ matrix rather than a $2^N \times 2^N$ one. 

Kitaev-type exchange interactions are significant in spin-orbit coupled Mott insulators~\cite{Khaliullin_2010} such as the iridium oxide family A$_2$IrO$_3$ (A = Na, Li) and $\alpha-$RuCl$_3$~\cite{chaloupkaKitaevHeisenbergModelHoneycomb2010,rauSpinOrbitPhysicsGiving2016,plumbEnsuremathAlphaEnsuremath2014, takagiConceptRealizationKitaev2019}. 
This has resulted in a flurry of research in the search for an experimental realization of the Kitaev quantum spin liquid~\cite{banerjeeProximateKitaevQuantum2016, jansaObservationTwoTypes2018,kasaharaMajoranaQuantizationHalfinteger2018}. 
These materials exhibit additional interaction terms beyond the Kitaev exchange and show a rich behavior under an external magnetic field, which cannot be treated exactly within the fermionic description and typically requires a numerical analysis. 
Many numerical studies using various techniques such as exact diagonalization (ED), density-matrix renormalization group (DMRG), and tensor network (TN) methods have revealed new and exotic phases of the model beyond the perturbative regime~\cite{Hickey_Trebst_2019, Khaliullin_2010,Trebst_2011,Khaliullin_2013,Troyer_2014,Kee_2014,Tohyama_2015,Pollmann_2017,Andrzej_2017,Imada_2015,Motome_2021}. 
Effective field theory techniques can also provide valuable insight into the behavior in a magnetic field~\cite{zhangTheoryKitaevModel2022}. 

{Quantum computers offer an exciting new framework for simulating quantum many-body systems.}
There are a number of efforts exploring simulation of the Kitaev model on quantum computers~\cite{Andy_2021,Kyriienko_2021, lex_2021, Schmied_2011}. Connections of the Kitaev honeycomb model to quantum error correction have also been explored previously \cite{Suchara_2011,PhysRevA.81.032301,Wootton_2015,Lee_2017}, and more recently, it was shown that one could obtain a logical subspace out of an empty subsystem code defined on the honeycomb lattice through particular measurement schedules \cite{Hastings2021dynamically,https://doi.org/10.48550/arxiv.2110.09545,Gidney2021faulttolerant,https://doi.org/10.48550/arxiv.2202.11829,https://doi.org/10.48550/arxiv.2203.11137}.
In this paper we focus on variational eigenstate preparation in Kitaev models with and without integrability breaking terms. To inject information about the exact solvability of the model in a certain regime, we here propose using a fermionic description to simulate the model on a quantum computer. As we show, in certain situations this allows reducing the number of required qubits by half compared to VQE approaches that are formulated within the spin description~\cite{Andy_2021, Kyriienko_2021}. One interesting application of our method, on which we comment in Sec.~\ref{sec:anyon}, is the simulation of non-Abelian anyons on quantum computers. While the fermionic description can reduce the required qubit number, a drawback of simulating fermions on quantum computers is the need for a mapping from the fermionic Hilbert space to that of qubits, which necessitates deeper quantum circuits~\cite{Seeley_Richard_Love_2012,Jordan_Wigner_1928,Kitaev_2002, Cirac_2005,Tavernelli_2016, Troyer_2016}. 
This added circuit depth can make the quantum circuits challenging to run on current noisy intermediate-scale quantum (NISQ)~\cite{Preskill2018quantumcomputingin} hardware. Whether the qubit reduction still offers an advantage on NISQ hardware should be explored in future work. 

\begin{figure*}[t]
    \centering 
    \subfloat[Honeycomb lattice]{ \includegraphics[trim = 0 -8 100 0, clip, scale=0.85]{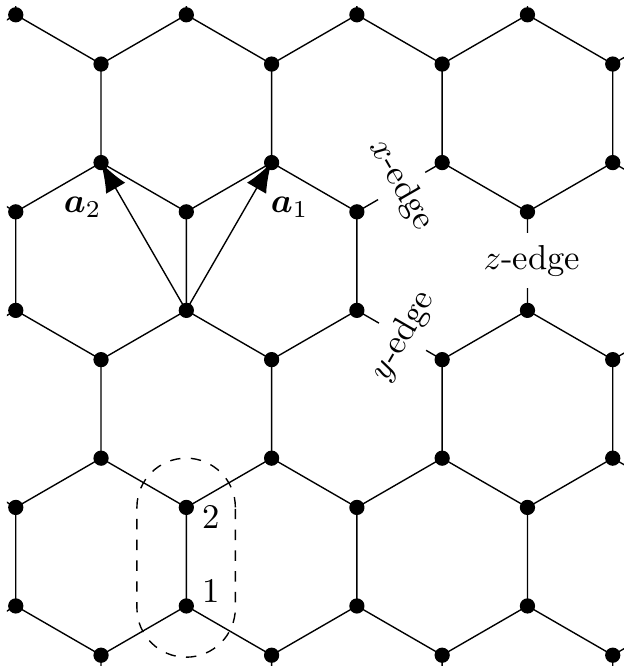} }
    \qquad
    \subfloat[Square-octagon lattice]{
        \includegraphics[trim = 0 0 28 28, clip, scale=0.9]{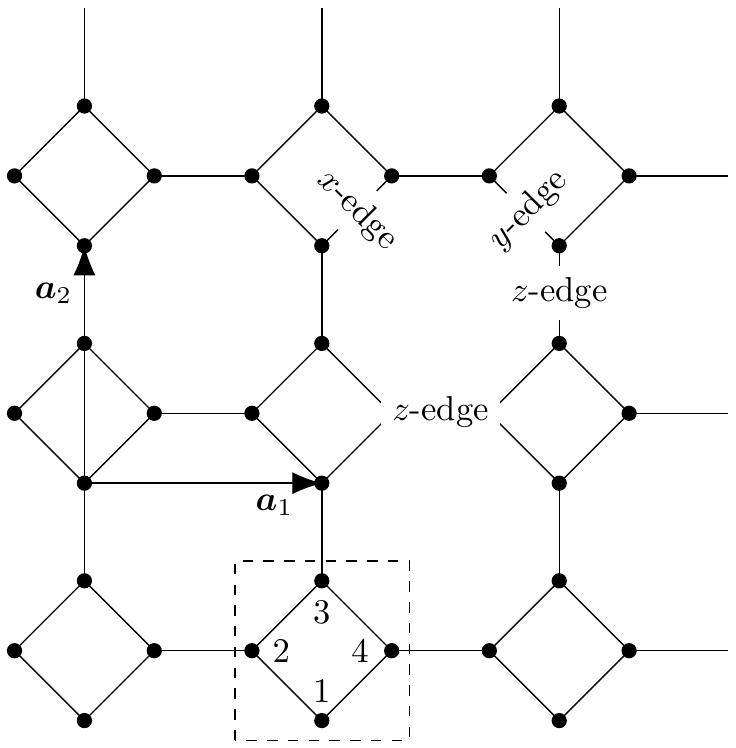}} 
    \begin{minipage}[b]{0.3\textwidth}
        {\includegraphics[trim = 0 0 0 0, clip, scale=0.75]{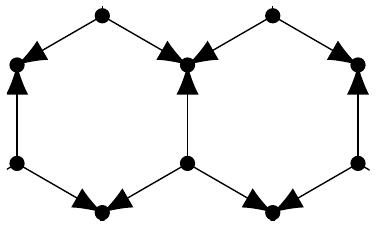} }
        
        {\includegraphics[trim = 0 0 0 0, clip, scale=0.75]{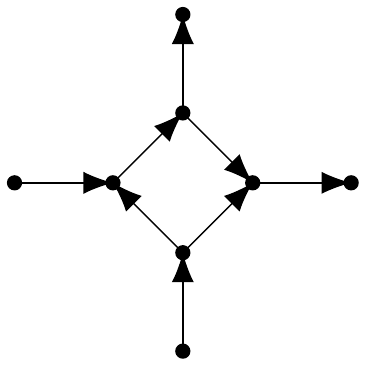}}    
        
        \subfloat[Standard gauge]{
        \raisebox{15pt}{
        $u^{\text{std}}_{ji} = 
        \begin{cases}
            +1 \quad j \tikz[baseline=-0.23em]{\draw[{Latex[length=2mm]}-] (0,0) -- (0.6,0) ;} i \\
            -1 \quad j \tikz[baseline=-0.23em]{\draw[-{Latex[length=2mm]}] (0,0) -- (0.6,0) ;} i
        \end{cases}$} }
    \end{minipage}
    \caption{(a) Honeycomb lattice and (b) square-octagon lattice with definitions of primitive unit cell vectors $\bm{a}_1$ and $\bm{a}_2$ and edge labels. The dashed lines enclose the smallest unit cell of the lattices, which include the labels of the basis sites. Panel (c) shows the standard gauge choice of the bond variables $u_{ji}$ for both lattices.}
    \label{fig:def_lattices}
\end{figure*}

We make use of the variational quantum eigensolver (VQE)~\cite{Peruzzo2014,cerezo2020variational, bhartiNoisyIntermediatescaleQuantum2022}, a hybrid algorithm (i.e., one using both classical and quantum computers) with significant potential for successful implementation on NISQ devices~\cite{Peruzzo2014,OMalley_PRX_2016,Kandala_Nature_2017,Hempel-PRX_2018,McCaskey_npjQI_2019, hartree_fock_google_2020, mukherjeeComparativeStudyAdaptive2022}. 
A VQE algorithm uses a quantum computer to prepare a variational ansatz state, defined using a parameterized quantum circuit, and then measures its energy (i.e., the expectation value of the Hamiltonian in that state).
A classical computer is then used to find the optimal set of variational parameters that produces the lowest possible energy expectation value. 
On a classical computer, preparing the state and calculating the energy expectation value are computationally expensive, so handing these steps over to a quantum computer may offer an effective speed-up. 
VQE algorithms offer shorter circuits when compared to other methods like adiabatic real-time evolution~\cite{Farhi00}, quantum imaginary time evolution~\cite{motta_2019, gomesEfficientStepMergedQuantum2020a}, or phase estimation~\cite{nielsenQuantumComputationQuantum2010}, and thus are {viewed as being well-suited for execution} on NISQ devices~\cite{Peruzzo2014, McClean_2016, cerezo_2021, bhartiNoisyIntermediatescaleQuantum2022}. 

The Kitaev spin model with its bond-dependent interactions can be defined on any trivalent graph, and in this work we focus on the honeycomb and square-octagon lattices. 
The exact solution of the model relies on a mapping to a model of Majorana fermions coupled to a $\mathbb Z_2$ lattice gauge field. 
In this work, we use classical hardware to perform VQE simulations of the Kitaev model in the presence of two kinds of additional Hamiltonian terms. 
First, {there} are 3-spin interaction terms that do not mix different gauge sectors of the model. 
These terms {allow} for the calculation to be restricted to a single gauge sector and {lead} to a twofold reduction {in} the number of qubits.
We also consider external magnetic fields in the $x,y$, and $z$-directions, which mix different gauge sectors together, and in this case we include the full Hilbert space in the calculation. Then, twice as many qubits as spins are needed in the simulation.

The rest of the paper is organized as follows. In Sec.~\ref{sec:review_of_model} we give a brief review of certain aspects {of} the Kitaev model that are important {for our analysis}.
Then in Sec.~\ref{sec:fg_part} we discuss the calculation when restricted to a single gauge sector, and discuss the application of realizing non-Abelian anyons on quantum computers. 
Finally in Sec.~\ref{sec:mg_part} we discuss how to extend the calculation to include all gauge sectors of the model. 

\section{Kitaev model and fermionic formulation}\label{sec:review_of_model}
\subsection{Kitaev spin Hamiltonian}
A trivalent lattice is one in which every site is connected to three other sites---a condition satisfied, for example, by both the honeycomb and square-octagon lattices as shown in Fig.~\ref{fig:def_lattices}.
Throughout the text, we reserve the labels $i, j, k, \dots$ for the lattice sites. 
The trivalence of the lattice allows for the edges to be be split into three disjoint sets, which will be referred to as $x, y,$ and $z$-edges. 
The designation of $x,y,$ and $z$-edges for both the honeycomb and square-octagon lattices is shown in Fig.~\ref{fig:def_lattices}.
The Hamiltonian of the Kitaev model is given as, 
\begin{align}\label{eq:pure_kitaev_model}
    H = -\sum_{\alpha=x,y,z} J_\alpha \sum_{\alpha-\text{edges}} \sigma^{\alpha}_i \sigma^{\alpha}_j 
\end{align}
where $\sigma_i^{\alpha}$ are Pauli operators at site $i$ and $\alpha = x,y,z$. The summation over edges counts every lattice bond of type $\alpha$ once.  
Explicitly, on the honeycomb lattice, which has two basis sites $\tau = 1, 2$ per unit cell, it can be written as $H= - \sum_{\alpha} J_\alpha \sum_{\bm{r}_i} \sigma^\alpha_{\bm{r}_i, 1} \sigma^\alpha_{\bm{r}_i + \boldsymbol{\delta}_{\alpha}, 2}$, where $\bm{r}_i = i_1 \bm{a}_1 + i_2 \bm{a}_2$, and $\boldsymbol{\delta}_{x} = - \bm{a}_1$, $\boldsymbol{\delta}_{y} = - \bm{a}_2$, $\boldsymbol{\delta}_{z} = 0$. 
The unit cell vectors $\bm{a}_i$ are shown in Fig.~\ref{fig:def_lattices}(a). 
The square-octagon lattice has four basis sites per unit cell, $\tau = 1, 2, 3, 4$, and its Hamiltonian reads explicitly as
$H= - \sum_{\bm{r}_i} J_x  \bigl( \sigma^x_{\bm{r}_i,2} \sigma^x_{\bm{r}_i,3} + \sigma^x_{\bm{r}_i,4} \sigma^x_{\bm{r}_i,1}\bigr) + J_y  \bigl( \sigma^y_{\bm{r}_i,1} \sigma^y_{\bm{r}_i,2} + \sigma^y_{\bm{r}_i,3} \sigma^y_{\bm{r}_i,4}\bigr) + J_z  \bigl( \sigma^z_{\bm{r}_i,4} \sigma^z_{\bm{r}_i + \bm{a}_1,2} + \sigma^z_{\bm{r}_i,3} \sigma^z_{\bm{r}_i + \bm{a}_2,1}\bigr)$. The basis labels $\tau$ and unit cell vectors $\bm{a}_i$ are shown in Fig.~\ref{fig:def_lattices}(b).

The Kitaev model~\eqref{eq:pure_kitaev_model} has a conserved quantity associated with each plaquette $p$. 
For the honeycomb lattice there is only one kind of plaquette, and the conserved quantity $[W_p^{(6)}, H] = 0$ takes the form
\begin{align}\label{eq:w_honey_plaquette}
    W^{(6)}_p =  \sigma_1^x \sigma_2^y \sigma_3^z \sigma_4^x \sigma_5^y \sigma_6^z , && \ \   \raisebox{-37pt}{ \tikz[line join = round]{ \draw[black] (30:1) -- (90:1) -- (150:1) -- (210:1) -- (270:1) -- (330:1) --cycle; \draw[fill] (210:1) circle (2pt) -- (210:1.5) node[pos = -0.75] { \footnotesize $1$} (150:1) circle (2pt) -- (150:1.5) node[pos = -0.75] { \footnotesize $2$} (90:1) circle (2pt) -- (90:1.5) node[pos = -0.75] { \footnotesize $3$} (30:1) circle (2pt) -- (30:1.5) node[pos = -0.75] { \footnotesize $4$} (330:1) circle (2pt) -- (330:1.5) node[pos = -0.75] { \footnotesize $5$} (270:1) circle (2pt) -- (270:1.5) node[pos = -0.75] { \footnotesize $6$} ; }}.
\end{align}
For the square-octagon lattice there are two kinds of plaquettes, giving rise to two distinct plaquette operators ($[W^{(4)}_p, H] = [W^{(8)}_p, H] = 0$):
\begin{align}\label{eq:w_sq_oct_plaquette}
    &W^{(4)}_p =  \sigma_1^z \sigma_2^z \sigma_3^z \sigma_4^z , && \ \raisebox{-27pt}{ \tikz[line join = round]{ \draw[black] (45:1) -- (135:1) -- (225:1) -- (315:1) --cycle; 
    \draw[fill] (225:1) circle (2pt) -- (225:1.5) node[pos = -0.75] { \footnotesize $1$} (135:1) circle (2pt) -- (135:1.5) node[pos = -0.75] { \footnotesize $2$} (45:1) circle (2pt) -- (45:1.5) node[pos = -0.75] { \footnotesize $3$} (315:1) circle (2pt) -- (315:1.5) node[pos = -0.75] { \footnotesize $4$} ; 
    }} \\ 
    &W^{(8)}_p =  \sigma_1^x \sigma_2^y \sigma_3^y \sigma_4^x \sigma_5^x \sigma_6^y  \sigma_7^y \sigma_8^x , &&  \raisebox{-37pt}{ \tikz[line join = round]{ \draw[black] (22.5:1) -- (67.5:1) -- (112.5:1) -- (157.5:1) -- (202.5:1) -- (247.5:1) -- (292.5:1) -- (337.5:1) --cycle; 
    \node at (202.5:0.75) {\footnotesize $1$} ;
    \node at (157.5:0.75) {\footnotesize $2$} ;
    \node at (112.5:0.75) {\footnotesize $3$} ;
    \node at (67.5:0.75) {\footnotesize $4$} ;
    \node at (22.5:0.75) {\footnotesize $5$} ;
    \node at (337.5:0.75) {\footnotesize $6$} ;
    \node at (292.5:0.75) {\footnotesize $7$} ;
    \node at (247.5:0.75) {\footnotesize $8$} ;
    \draw[fill] (202.5:1) circle (2pt) -- ++(225:0.5) (157.5:1) circle (2pt) -- ++(135:0.5) (112.5:1) circle (2pt) -- ++(135:0.5) (67.5:1) circle (2pt) -- ++(45:0.5) (22.5:1) circle (2pt) -- ++(45:0.5) (337.5:1) circle (2pt) -- ++(-45:0.5)  (292.5:1) circle (2pt) -- ++(-45:0.5) (247.5:1) circle (2pt) -- ++(225:0.5) ;
     }}  
\end{align}
Note that all $W_p$ have eigenvalues of $\pm 1$ since $W_p^2 = 1$. 
It is useful to decompose the Hilbert space {into blocks labeled by the eigenvalues of $W_p$, i.e., $\mathcal L = \bigoplus_w \mathcal L_w$, where $\mathcal L$ is the full Hilbert space and $\mathcal L_w$ {denotes the eigenspace corresponding to a particular combination $w$ of eigenvalues of the various $W_p$ operators}. 
A theorem by Lieb~\cite{Lieb_1994} tells us that the ground state belongs to the sector with all $W_p = +1$. 
This sector is referred to as the vortex-free sector. 

\subsection{Representing spins using Majorana fermions}
\label{sec:spin_to_majorana}
The Hilbert space $\mathcal L$ of the lattice is the tensor product of the Hilbert spaces $\mathcal L_i$ of each spin, $\mathcal L = \bigotimes_i \mathcal L_i$. 
We seek a representation of the local spin Hilbert space using two fermionic degrees of freedom at each site, or four Majorana fermions. 
This fermionic Hilbert space is labeled as $\tilde{\mathcal L}_i$, with $b^{x,y,z}_i$, and $c_i$ being the four Majorana fermions at each site.  
These Majorana operators obey the algebra 
\begin{align}
    \{b^\alpha_i, b^\beta_j\} = 2 \delta_{ij} \delta_{\alpha \beta}, \indent \{c_i,c_j\}=2\delta_{ij} ,\indent  \{b^\alpha_i, c_j\} =  0. 
\end{align}
The two-dimensional Hilbert space $\mathcal{L}_i$ is the physical subspace of the four-dimensional Hilbert space $\tilde{\mathcal{L}}_i$. A physical state $\ket{\psi_{\text{phys}}}_i \in {\mathcal{L}}_i$ is defined such that 
\begin{align}
    D_i \ket{\psi_{\text{phys}}}_i = \ket{\psi_{\text{phys}}}_i, && D_i = b^x_ib^y_ib^z_ic_i.
\end{align}
The operator $D_i$ acts on the physical subspace as an identity, and since we are only interested in this subspace it should be noted that two operators differing only by factors of $D_i$ are identified in this treatment.
Further, given any $\ket{\psi}_i \in \tilde{\mathcal{L}}_i$, the physical part of this state can be extracted as follows: 
\begin{align}
    \ket{\psi_{\text{phys}}}_i = \frac{1}{2}(1 + D_i) \ket{\psi}_i.
\end{align}
Thus, the operator $(1 + D_i)/2$ is the local projection operator onto the physical subspace. The full projector can be written as
\begin{align}\label{eq:full_projection_op}
    \mathcal{P} = \prod_i \mathcal P_i = \prod_i \frac{1 + D_i}{2}.
\end{align}

In terms of the Majorana fermions, the Pauli operators take the following form: 
\begin{align}
    \sigma^x_i = ib^x_ic_i, && \sigma^y_i = ib^y_ic_i && \sigma^z_i = ib^z_ic_i.
\end{align}
Using this representation of the Pauli operators, the Kitaev model can be written as 
\begin{align}\label{eq:kitaev_majorana_hamiltonian}
    H = \sum_\alpha J_\alpha \sum_{\alpha-\text{edges}} i \hat u_{ij}c_i c_j,
\end{align}
where 
\begin{align}
    \hat u_{ij} = ib^{\alpha}_i b^{\alpha}_j, && \hat u_{ij}^2 = 1, && \hat u_{ij} = -\hat u_{ji}.
\end{align}
Note that the eigenvalues $u_{ij}$ of {$\hat u_{ij}$} are $u_{ij} = \pm 1$ since {$\hat u_{ij}^2 = 1$}. 
The operator $\hat u_{ij}$ can be interpreted as a $\mathbb Z_2$ gauge field that couples to the itinerant Majorana fermions $c_i$. For this reason, we will sometimes refer to the $c_i$ Majorana fermions as ``matter" fermions, to distinguish them from the ``bond" fermions $b^\alpha_i$.
The operator $D_i$ anticommutes with $\hat u_{ij}$ and therefore can be interpreted as implementing a gauge transformation that flips the value of $u_{ij}$. 

\subsection{The single particle transformation diagonalizing the Kitaev model}\label{sec:reduce_quad}
As noted by Kitaev, the operators $\hat u_{ij}$ commute with all terms in the Hamiltonian, so the eigenvalues $u_{ij}=\pm 1$ are conserved quantities of the model. Thus, it is useful to write  
\begin{align}
    \tilde{\mc L} = \bigoplus_u \tilde{\mc L}_u,
\end{align}
where $\tilde{\mc L}_u$ is the subspace with all $u_{ij}$ specified. 
The conserved quantities $W_p$ can be expressed in terms of $u_{ij}$ as follows:
\begin{align}
\label{eq:Wp}
    &W^{(6)}_p = \prod_{i \in p} u_{i+1 , i}, \nonumber \\ 
    &W^{(4)}_p = - \prod_{i \in p} u_{i+1 , i},  && W^{(8)}_p = - \prod_{i \in p} u_{i+1 , i}. 
\end{align}
Thus each subspace $\tilde{\mathcal L}_u$ corresponds to a certain configuration of $W_p$. We will sometimes refer to $\tilde{\mathcal L}_u$ as a ``gauge sector," i.e. a sector of the full Hilbert space whose gauge has been fixed by a choice of the eigenvalues $u_{ij}$.

As noted previously, the ground state belongs to the vortex-free configuration. 
There are many configurations of $u_{ij}$ that give the vortex-free configuration. 
Fig.~\ref{fig:def_lattices} (c) and (d)  define our choice of a ``standard configuration" $u^{\text{std}}_{ij}$ for both the honeycomb and square-octagon lattices, which is a simple choice of gauge such that all $W_p = 1$.

In the subspace $\tilde{\mc L}_{u}$, the Hamiltonian in Eq.~(\ref{eq:kitaev_majorana_hamiltonian}) takes the following quadratic form: 
\begin{align} \label{eq:qudratic_hamiltonians}
    H = \frac{i}{2}\sum_{i,j = 1}^{N} K_{ij} c_ic_j,
\end{align} 
where the matrix $K = u_{ij}$ when $i$ and $j$ make an edge and $K_{ij}=0$ otherwise. Note that $K_{ji} = - K_{ij}$. In order to diagonalize a Hamiltonian of this form we need to find a matrix $R \in O(N)$ such that 
\begin{align}\label{eq:block_diagonal_quad_ham}
    RKR^{T} = \bigoplus_{n=1}^{N/2} 
    \begin{bmatrix} 
        0 &   \varepsilon_n \\ 
        -\varepsilon_n & 0 
    \end{bmatrix}, && \varepsilon_n \ge 0. 
\end{align}
This transformation can be achieved by a unitary matrix $U$, 
\begin{align}\label{eq:majorana_R_transformation}
    U^{-1} c_i U = \sum_j R_{ji} c_j. 
\end{align}
such that, 
\begin{align}
    U^{-1} H U &= \frac{i}{2} \sum_{i,j = 1}^N \[R K R^T\]_{ij} c_i c_j \\ 
    & = i \sum_{i = 1}^{N/2} \varepsilon_i c_{2i} c_{2i + 1}.
\end{align}
Note that the operator $U^{-1}HU$ is different from the operator $H$, as $U$ does not commute with $H$. 
To read off the spectrum, it is useful to pair the Majorana fermions into complex fermions. 
How the Majorana fermions are paired into complex fermions is a matter of basis choice. 
Here we choose to couple the Majorana fermions inside the same unit cell together. 
For the honeycomb lattice, the $1$ sublattice is paired with the $2$ sublattice, and for the square-octagon lattice, the $1$ sublattice is paired with the $2$ sublattice, and the $3$ sublattice is paired with the $4$ sublattice. 
Such a choice of basis can be written in the following way,
\begin{align}
    c_{2j} = a_j + a_j^\dagger, \quad c_{2j + 1} =  \frac{1}{i}(a_j - a^\dagger_j), \label{eq:c_a_mapping} \\ 
    U^{-1} H U = \sum_{i=1}^{N/2} 2 \varepsilon_i \left(a^\dagger_i a_i - \frac{1}{2}\right). 
\end{align}
The ground state of $H$ can be written as $U \ket{\psi_0}$, where
\begin{align}\label{eq:init_state}
    a_i \ket{\psi_0} = 0, \ \text{for all } a_i.
\end{align} 
The action of the Hamiltonian on $U \ket{\psi_0}$ is found to be
\begin{align}
    H U \ket{\psi_0} = E_0 U\ket{\psi_0}, \\
    E_0 = -\sum_{i=1}^{N/2} \varepsilon_i.  
\end{align}

In designing our VQE ansatz it will be crucial to know what form the operator $U$ takes. 
A general $SO(N)$ transformation can be applied using $\exp\left[\sum_{ij} \theta_{ij} c_i c_j \right]$, which acts on a Majorana operator $c_i$ as
\begin{align} \label{eq:su2_majorana_transformation}
    \exp\left[-\sum_{ij} \theta_{ij} c_i c_j \right] c_i \exp\left[\sum_{ij} \theta_{ij} c_i c_j \right]  = [e^{\bm \theta}]_{ji} c_j. 
\end{align}
Even though any antisymmetric matrix can be brought to the block diagonal form in Eq.~(\ref{eq:block_diagonal_quad_ham}) by an $SO(N)$ transformation, to ensure that the upper-right element of each block is a positive number (as required) we need to be allowed $O(N)$ transformations. 
This can be seen by noting that the operation of exchanging the off-diagonal elements of a $2 \times 2$ matrix (\textit{i.e.}, $\sigma^x$) is an operation with determinant $-1$.
Thus, we might need to attach a local ``particle-hole" transformation to $ \exp\[\sum_{ij} \theta_{ij} c_i c_j \]$ to make sure all $\varepsilon_i \geq 0$.
Note that this operation would only be needed if an odd number of the $2 \times 2$ blocks need such an operation.
For example, switching the off-diagonal parts of two of these $2 \times 2$ blocks can be done by a $\sigma^x \oplus \sigma^x$ which has a determinant of $+1$, and is expressible by $ \exp\[\sum_{ij} \theta_{ij} c_i c_j \]$.
In short, it is just the determinant of the transformation that we need to worry about. 

This shows that the pure Kitaev model in Eq.~(\ref{eq:pure_kitaev_model}) is exactly solvable. 
In the next section we discuss several additional terms of interest that spoil the exact solvability of the model. 
The form of the exact solution will still be a useful guide when choosing the form of the variational ansatz in the VQE calculation.
If we always include $U$ as a part of the ansatz we make sure the algorithm can exactly reproduce the ground state in the exactly solvable limit, where the model is quadratic in terms of fermion operators.
We will also add more terms to the ansatz in order to better approximate the ground state in the presence of interactions as we discuss next.

\subsection{Added interactions}
The terms that can be added to the pure Kitaev Hamiltonian fall into two classes: the first class contains terms that do not mix different flux sectors, and the second class contains terms that do. 
Here we consider both kinds of terms. 
This distinction is useful because it informs us how the model will be simulated on the quantum computer. 
For terms of the first kind we only need to simulate a single gauge sector of the model, which is a much smaller Hilbert space than that of the original spin Hilbert space, and does reduce the number of qubits needed for the calculation.

Terms that do not mix different gauge sectors are of the following form,
\begin{align}
    V = -\sum_{(i,j,k;l)} &\[\kappa ( \sigma^x_i \sigma^y_j \sigma^z_l + \sigma^x_i \sigma^y_l \sigma^z_k + \sigma^x_l \sigma^y_j \sigma^z_k) \right . \nonumber \\
    &\left .  + \kappa_{\text{int}} \sigma^x_i \sigma^y_j \sigma^z_k \],
    \label{eq:three_spin_terms}
\end{align}
where $(i,j,k;l)$, refers to the $i,j$, and $k$-th sites connected to the $l$-th site as shown in Fig.~\ref{fig:3-spin_terms}.
These terms show up at $3^{\text{rd}}$ order when treating an external magnetic field perturbatively. 
However, we will study the effects of these terms regardless of their origin 
and treat $\kappa$ and $\kappa_{\text{int}}$ as independent parameters. 

\begin{figure}
    \centering
    \includegraphics{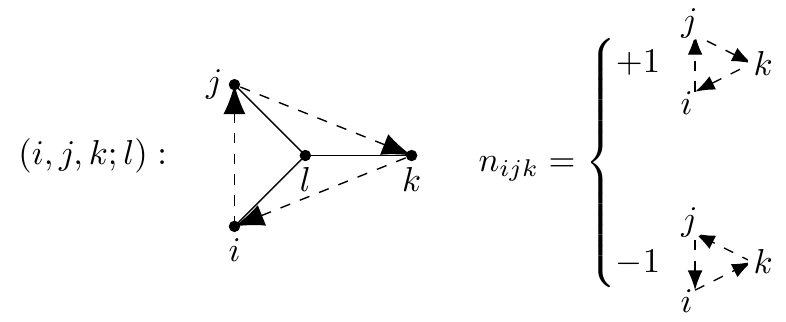}
    \caption{The definition of the 3-spin terms in Eq.~\eqref{eq:three_spin_terms} that are gauge diagonal. Such terms appear when performing perturbation theory in a small magnetic field~\cite{Kitaev_2006}. The quantity $n_{ijk}$ appears in the fermionic representation of these terms in Eq.~\eqref{eq:kappa_terms}.}
    \label{fig:3-spin_terms}
\end{figure}

The $\kappa$ and $\kappa_{\text{int}}$ terms map to very different looking terms on the fermionic side~\cite{Kitaev_2006}, 
\begin{align}
    V = \sum_{(i,j,k;l)}  n_{ijk} &\[  i \kappa( u_{il}u_{lj}\,   c_j c_i +  u_{ik}u_{li}\,   c_i c_k +  u_{jl}u_{lk}\,   c_k c_j ) \right. \nonumber \\
    + & \ \left.  \kappa_{\text{int}}\,  u_{il} u_{jl} u_{kl}\,  c_i c_j c_k c_l\],
    \label{eq:kappa_terms}
\end{align}
where $n_{ijk}$ is defined as in Fig.~\ref{fig:3-spin_terms}. The $\kappa$ terms correspond to second-neighbor hopping terms of $c_i$ fermions and preserve the exact solvability of the model. They are important as they can drive the system into a topologically ordered state. 
The $\kappa_{\text{int}}$ terms, on the other hand, are four-fermion terms (hence the subscript as a reminder that these terms add interactions to the model), and thus spoil the exact solvability of the model. Their effects are less well studied in the literature. 
Later we discuss one aspect in which these terms can be interesting and useful in stabilizing Majorana zero modes localized at vortices. 

As an example of terms that mix different gauge sectors, we will consider a uniform external magnetic field, 
\begin{align}\label{eq:spin_ex_mag}
    H_{\text{mag}} = - \sum_i \[ h_x \sigma^x_i + h_y \sigma^y_i +h_z \sigma^z_i\]. 
\end{align}
In the language of the fermionic degrees of freedom this can be written as, 
\begin{align}\label{eq:fermion_magnetic_field_hamiltonian}
    H_{\text{mag}} = -i \sum_i \[ h_x b^x_i c_i + h_y b^y_ic_i +h_z b^z_ic_i \]. 
\end{align}
When simulating the Kitaev model in an external magnetic field, we therefore must include all gauge sectors in the calculation. 

\section{VQE}\label{sec:main_work}
A VQE algorithm contains four parts: first, one prepares an initial state $\ket{\psi_0}$, which is typically a state that can be easily prepared on the quantum device. 
Second, one applies a parameterized unitary (or quantum circuit) $U(\bm \theta)$ with variational parameters $\bm \theta$ to the initial state to prepare the ansatz wavefunction $\ket{\psi(\bm \theta)} = U(\bm \theta) \ket{\psi_0}$. 
The third step is to measure a cost function $C(\bm \theta)$, which is a sum of observables that are being measured in the variational state $\ket{\psi(\bm \theta)}$. 
To prepare the ground state of a system, the cost function is usually taken to be the energy expectation value $C(\bm \theta) = \ev*{H}{\psi(\bm \theta)}$. 
However, as we will discuss in the dynamical-gauge VQE section, it can be useful to use a slightly modified cost function. 
Finally, the fourth step is the classical optimization over the set of parameters $\bm \theta$ so as to minimize $C(\bm \theta)$. 
This involves frequent evaluations of the cost function that follow the first three steps. 
A VQE algorithm is designed such that the first three steps are carried out on a quantum computer while the fourth is done on a classical computer. 

\subsection{Fixed-gauge VQE}\label{sec:fg_part}
\subsubsection{Matter sector initial state}
Even though the model is most conveniently expressed in terms of Majorana fermions, for the sake of simulating the system on a quantum computer, we need to group the Majorana fermions into pairs of complex fermions in order to map the problem onto qubits. 
We already discussed how we choose to group the the $c_i$ Majoranas into the complex fermions $a_i$ in Eq.~(\ref{eq:c_a_mapping}), namely, 
\begin{align}
    a_i = \frac{1}{2} (c_{2i} + i c_{2i + 1}), && a^\dagger_i = \frac{1}{2} (c_{2i} - i c_{2i + 1}).
\end{align}
Note that the label $i$ in $a_i$ refers to a unit cell location $\bm r_i$ for the honeycomb model, and is a composite index that labels both unit cell location and a Majorana pair $(1,2)$ or $(3,4)$ for the square-octagon model. 
For the purpose of finding the ground state we choose an initial state in the vortex-free sector of the  Hilbert space, where the plaquette operators $W_p = 1$ for all $p$.  
Though it should be mentioned that we could also choose any other vortex configuration. 
This will be useful later when discussing the possible application of realizing non-Abelian anyons. 
Further, we also choose the initial state of the system to be annihilated by all $a_i$, as defined in Eq.~(\ref{eq:init_state}).
\begin{align}\label{eq:init_state_def}
    \ket{\psi_0} \in \tilde{\mc L}_{u^{std}}, && a_i \ket{\psi_0} = 0 \ \  \forall a_i. 
\end{align}
After a Jordan-Wigner transformation, the details of which are discussed in Appendix~\ref{app:JW}, this initial state would simply correspond to the $\ket{0}$ state on the quantum computer, i.e., the ``all-0" state in the $Z$ eigenbasis.

\subsubsection{Variational ansatz}
\label{sec:fg_ansatz}
When performing VQE in the fixed-gauge subspace we use an ansatz of the following form: 
\begin{align}
    \ket*{\psi(\bm \theta)} &= \exp\[\sum_{ijkl} \theta^b_{ijkl} c_i c_j c_k c_l\] \exp\[\sum_{ij}\theta^a_{ij} c_i c_j\] \ket*{\psi_0} \nonumber \\ 
    & \equiv U^b(\bm \theta^b) U^a(\bm \theta^a) \ket{\psi_0} \equiv U(\bm \theta)\ket{\psi_0},
\end{align}
with both $\bm \theta^a$ and $\bm \theta^b$ being anti-symmetric under the exchange of any two indices, and having all components being real.
This form of the ansatz is motivated by the Hamiltonian variational ansatz successfully used in quantum chemistry and many-body problems~\cite{weckerProgressPracticalQuantum2015,Wiersema20}. 
It contains a unitary single-particle transformation term $U^a$, which can diagonalize the single-particle sector in the exactly solvable limit, and an interaction term $U^b$ that can account for additional correlations created by four-fermion interaction terms.
\begin{figure*}[t]
    \centering 
    \subfloat[$\kappa_{\text{int}} = 0$]{\includegraphics[trim = 0 0 0 0, clip, scale=0.65]{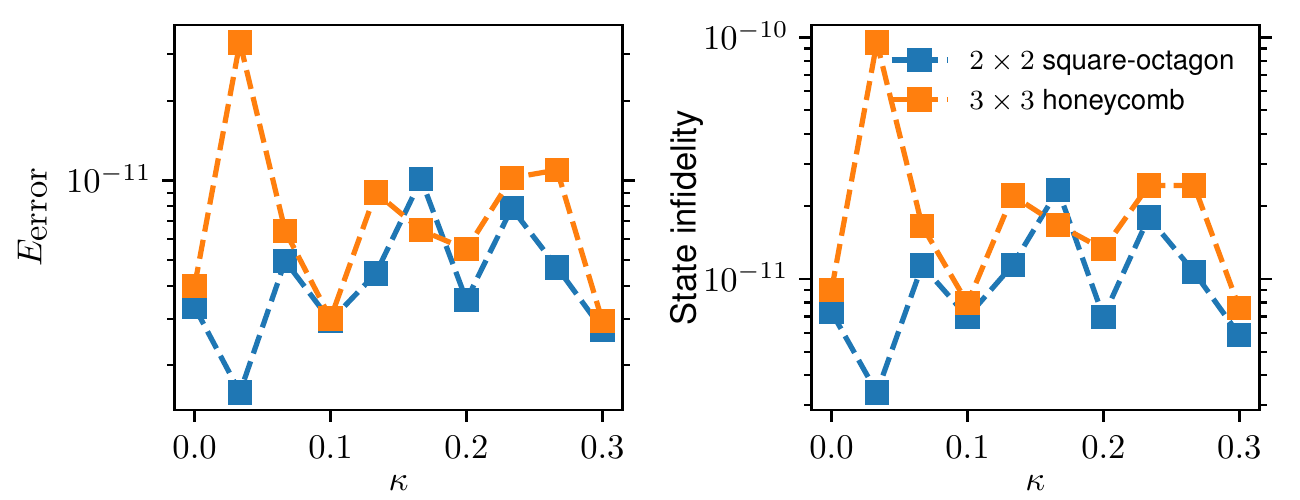} }
    \subfloat[$\kappa_{\text{int}} = \kappa$]{\includegraphics[trim = 0 0 0 0, clip, scale=0.65]{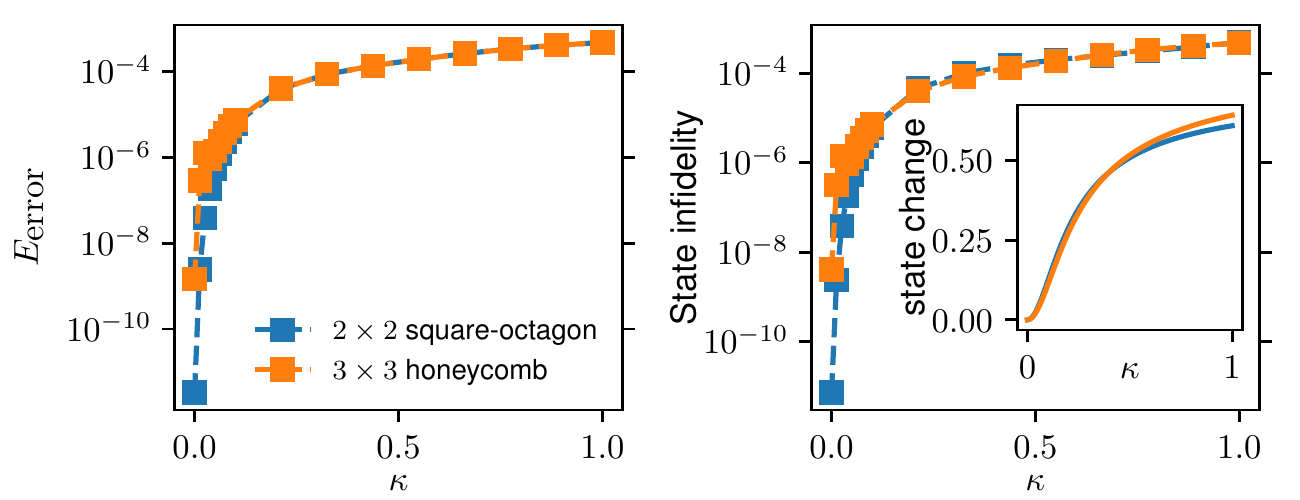} }
    \caption{VQE results using a statevector simulator for a fixed gauge configuration in $3 \times 3$ honeycomb model (orange) and $2 \times 2$ square-octagon model (blue). We show both the error in the ground state energy, $E_{\text{error}} =  \abs{(E_{\text{VQE}} - E_{\text{exact}})/E_{\text{exact}} }$, and the infidelity with respect to the exact ground state $1 - |\braket{\psi_{\text{exact}}}{\psi_{\text{VQE}}}|^2$. Panel (a) is for the exactly solvable quadratic model with and without second-neighbor hopping terms $\kappa$, while panel (b) includes fermion interactions $\kappa_\text{int} = \kappa$. In panel (b), we use the full ansatz defined in Eq.~\eqref{eq:ansatz_full} and in panel (a) we only include the $\bm \theta^a$ parameters. 
    Results in panel (a) demonstrate that our ansatz can reproduce the ground state energy in the exactly solvable model to arbitrary precision set by the error tolerance of the classical optimizer. We are effectively constructing the unitary $U$ that diagonalizes $K_{ij}$ variationally (see Sec.~\ref{sec:reduce_quad}). Results in panel (b) show the error and infidelity increase in the interacting model as a function of $\kappa_\text{int} = \kappa$, but saturate a low value of $\approx 10^{-3}$. This is at least one order of magnitude better $E_{\text{error}}$ when compared to other VQE methods in Ref.~\cite{Andy_2021}.
    Inset in (b) shows $1 - \abs{\braket{\psi(\kappa)}{\psi(0)}}^2$, where $\psi(\kappa)$ is the exact ground state for $\kappa = \kappa_{\text{int}}$. This is a quantification of how different the ground state at non-zero $\kappa=\kappa_{\text{int}}$ is compared to $\psi(0)$. Our method here has the advantage of cutting the required number of qubits by two compared to when simulating the model directly within the spin language.}
    \label{fig:fixed_gauge_vqe_results}
\end{figure*}

We focus on $U^a(\bm \theta^a)$ first.
We make this the first part of our ansatz since from our discussion in Sec.~\ref{sec:reduce_quad}, we know it should be capable of expressing the ground state of the pure Kitaev model. 
For a system with $N$ spins, there are $\frac{N(N-1)}{2}$ independent parameters in $\bm \theta^a$.
However, we are not interested in this full set of transformations. 
Rather we want to mod out the transformations that leave $\ket{\psi_0}$ invariant. 
We leave the details of such reduction of the ansatz to Appendix~\ref{app:relevant_ansatz}, and give the answer here in terms of the complex fermions $a_i$ defined in Eq.~(\ref{eq:c_a_mapping}),
\begin{align}
    &U^a(\bm \theta^a) \equiv \nonumber \\ 
    & \prod_{ij} \exp \[ i \theta^{a_1}_{ij}(a^\dagger_i a^\dagger_j  + a_ja_i) \] \exp \[ \theta^{a_2}_{ij}(a^\dagger_i a^\dagger_j - a_ja_i) \] 
\end{align}
Note that the number of complex fermions for a system described by $N$ Majorana fermions is $N/2$. Thus in total $\bm \theta^a$ contains $\frac{N}{2}\bigl(\frac{N}{2}-1 \bigr)$ independent parameters.

Recall the discussion below Eq.~(\ref{eq:su2_majorana_transformation}) about the determinant of the transformation needed to diagonalize the Hamiltonian. 
Since we fix our initial state in Eq.~(\ref{eq:init_state_def}), we might need to supplement $U^a(\bm \theta^a)$ with a local \emph{particle-hole} operation, in the cases where the ground state has different fermion parity to $\ket{\psi_0}$. 
This can easily be done by using $U^a(\bm \theta^a) c_1$  as the ansatz.
In all our simulations we compare the optimal energy resulting from using $U^a(\bm \theta)$ and $U^a(\bm \theta) c_1$, and report the one with lowest energy value. 

We also choose not to include all of the quartic terms in $U^b(\bm \theta^b)$ to simplify the circuits used.
Though it is not strictly the case, like before, that the dropped terms have no effect on the result, we found that only including the following terms offers the best performance in terms of computation time in our simulations:
\begin{align}
    &U^b(\bm \theta^b) \equiv \prod_{ijkl} \exp \[  i \theta^{b_1}_{ijkl}(a^\dagger_i a^\dagger_j a^\dagger_k a^\dagger_l   + a_l a_k a_j a_i) \] \nonumber \\
    &  \qquad \qquad    \times \exp \[ \theta^{b_2}_{ijkl}(a^\dagger_i a^\dagger_j a^\dagger_k a^\dagger_l - a_l a_k a_j a_i) \] .
\end{align}
With that being said, a more careful study of the effect of including the dropped terms might be in order, and we leave this for future work. 
The number of parameters contained in the above form of $U^b(\bm \theta^b)$ can be found to be $\frac{1}{4!}N \bigl(\frac{N}{2} - 1\bigr) \bigl( \frac{N}{2} - 2\bigr)\bigl(\frac{N}{2} - 3\bigr)$. Thus, the total number of parameters contained in $U(\bm \theta)$ is $\frac{N}{2} \bigl(\frac{N}{2} - 1\bigr)\[1 + \frac{1}{12}\bigl(\frac{N}{2} - 2\bigr) \bigl(\frac{N}{2} - 3\bigr)\]$. 
We discuss how to express this ansatz on a quantum computer in Appendix~\ref{app:JW}.

\subsubsection{Simulations and results}
In general, when restricting the Kitaev model with $N$ spins to a single gauge configuration, we end up with $N$ Majorana fermions $c_i$, one at each site $i$. 
This corresponds to $N/2$ complex fermions, and thus only $N/2$ qubits are needed for simulation. 
This is a substantial reduction compared to simulating the spins directly, which would require $N$ qubits. 
This reduction makes the fermionic formulations particularly attractive when considering additional terms in the Kitaev model that are ``gauge diagonal." 
Note that the model is no longer exactly solvable when quartic fermion interactions are present, which is where VQE calculations in the fermionic description will be most useful. 

We demonstrate the capabilities of the ansatz above using two geometries: a honeycomb lattice with $3 \times 3$ unit cells, and a square-octagon lattice with $2 \times 2$ unit cells.
These geometries have $18$ and $16$ spins respectively, and thus we only need $9$ and $8$ qubits for the VQE, which is a big advantage for our method. 
Periodic boundary conditions are applied in both cases. 
We set both models inside the gapless region of the phase diagram.
For the honeycomb lattice we set $\bm J = (J_x,J_y,J_z) = (1,1,1)$, and for the square-octagon lattice we set $\bm J = (1,1,\sqrt{2})$. 

In Fig.~\ref{fig:fixed_gauge_vqe_results}(a), we show the results of VQE simulation using a statevector simulator for the exactly solvable case $\kappa_{\text{int}} = 0$ as a function of second-neighbor hopping $\kappa$.
We plot the error in energy $E_{\text{error}} = \abs{(E_{\text{VQE}} - E_{\text{exact}}) / E_{\text{exact}}}$, and the state infidelity $1 - \abs{ \braket{\psi_{\text{exact}}}{\psi_{\text{VQE}}}}^2$, where $E_{\text{VQE}}$ and $\ket{\psi_{\text{VQE}}}$ are the optimal ground state energy and ground state obtained by VQE, while $E_{\text{exact}}$ and $\ket{\psi_{\text{exact}}}$ are the ground state energy and ground state obtained by exact diagonalization.
In this case, our ansatz can have arbitrary agreement with the exact ground state with the only bottleneck being the error tolerance we set for the classical optimizer. 

Moving away from the exactly solvable limit by including the four-fermion interaction term $\kappa_\text{int}$ in Eq.~\eqref{eq:kappa_terms}, Fig.~\ref{fig:fixed_gauge_vqe_results}(b) shows a sizeable decrease in the accuracy of the ansatz even though both $E_{\text{error}}$ and the state infidelity seemingly reach a plateau value of about $10^{-4}$, which is still quite a high accuracy. 
To put these numbers into perspective we compare our method to other VQE methods studied for the Kitaev model in Ref~\cite{Andy_2021}. 
Our method achieves at least one order of magnitude lower error in ground state energy when compared to all VQE methods studied in Ref~\cite{Andy_2021}. 
It is worth mentioning that even though the $3$-spin interaction terms considered here are different from the external magnetic field terms (see Eq.~(\ref{eq:spin_ex_mag})) considered in Ref.~\cite{Andy_2021}, comparisons are still instructive since the $3$-spin interaction terms are exactly the perturbative effects of the external magnetic field, especially considering that in Ref.~\cite{Andy_2021} small values of field $\bm h = (1,1,1)/\sqrt{3}$ where used. 

The inset in Fig.~\ref{fig:fixed_gauge_vqe_results} (b) shows the change of the ground state as a function of $\kappa$: $ 1 - \abs{\braket{\psi(\kappa)}{\psi(0)}}^2$, with $\ket{\psi(\kappa)}$ being the exact ground state as a function of $\kappa$. 
The fact that $ 1 - \abs{\braket{\psi(\kappa)}{\psi(0)}}^2 $ becomes much bigger than our state infidelity as $\kappa$ increases, demonstrates the excellent expressivity of our ansatz. 

Another advantage to motivating the variational ansatz using a fermionic description is the ability to simulate different vortex configurations of the model.
Even though Fig.~\ref{fig:fixed_gauge_vqe_results} shows results for the vortex free sector, adding a vortex would just correspond to a simple change on the Hamiltonian, $i.e.$ changing the corresponding signs of $u_{ij}$.
Other than that the VQE approach would behave in a very similar manner. 
Such a task would be very difficult for VQE using the spin language since higher vortex configurations would correspond to higher excited states which are challenging for a variational method to accurately simulate. 
Being able to simulate these vortex excitations has the possible exciting application of simulating non-Abelian anyons on quantum computers, as we discuss now. 

\subsubsection{Implications for quantum simulation of non-Abelian anyons}
\label{sec:anyon}
A potentially interesting application for our method is realizing non-Abelian anyons on quantum computers. 
Let us for now focus on the honeycomb lattice, though the square-octagon case is not substantially different. 
With $\bm J = (1,1,1)$ and $\kappa = \kappa_{\text{int}} = 0$, the model is gapless. Adding the $\kappa$ terms opens up a gap in the spectrum. One of the interesting features of the model in this region of the parameter space is that it hosts non-Abelian anyons~\cite{Kitaev_2006}. In particular, a vortex excitation of the model (i.e., a plaquette $p$ for which $W_p=-1$) will carry a Majorana zero mode. One can therefore imagine using VQE methods to prepare the ground state in the presence of some number of vortices. Then, by applying appropriate unitary transformations to this state (see, e.g., \cite{Xu11}), one could manipulate the vortices in order to ``braid" the attached Majorana zero modes, which is one route to realizing fault-tolerant Clifford operations~\cite{DasSarma15}. We discuss below some considerations that must be taken into account when contemplating such a scheme.

If we have an infinite system with two vortices that are very far from each other, we expect two degenerate ground states that have the same energy and different fermion parity. 
In both classical and quantum simulations, we only have access to finite systems and there is a limit to how far away the vortices can be from each other. As the Majorana modes get close to each other they can hybridize, leading to a small energy gap between the even- and odd-parity states. We henceforth refer to this energy scale as the ``ground-state splitting," to avoid confusion with the (larger) energy scale of the bulk gap, which is associated with creating a vortex excitation. In practice, it is desirable for this splitting to be as small as possible to suppress the accumulation of dynamical phases during braiding. 
The degree of closeness between the Majorana zero modes can be quantified by comparing the distance between the vortices to the correlation length $\xi$, defined as the localization length of the wavefunction of the Majorana bound state centered at the location of the vortex (which is inversely proportional to the bulk gap). 
To have robust Majorana modes on a quantum computer we have to be able to simulate systems whose sizes are of the order of $2\xi$ for periodic boundary conditions. 
Having open boundaries would not help since we will also need to require the Majorana modes to be away from the boundary.

It is thus desirable to make $\xi$ as small as possible, so that the vortices do not need to be very far apart during braiding.
For $\kappa \ll |\bm  J|$, one can show that $\xi \propto 1/\kappa$.
However, we also expect this behavior to change for large $\kappa$, since a theory with only $\kappa$ terms (without $J_x$, $J_y$ and $J_z$ terms) will be gapless, and thus has $\xi = \infty$.
We thus expect $\xi$ to have a minimum value as a function of $\kappa$.
This minimum value of $\xi$ is crucial since it puts a lower bound on the system sizes where we expect to observe the topological properties of the model  using only the $\kappa$ terms. 
This is one area where we find that including the $\kappa_{\text{int}}$ terms can be of some help, as we will now explain.

\begin{figure}[t]
    \centering
    \includegraphics[scale=0.7]{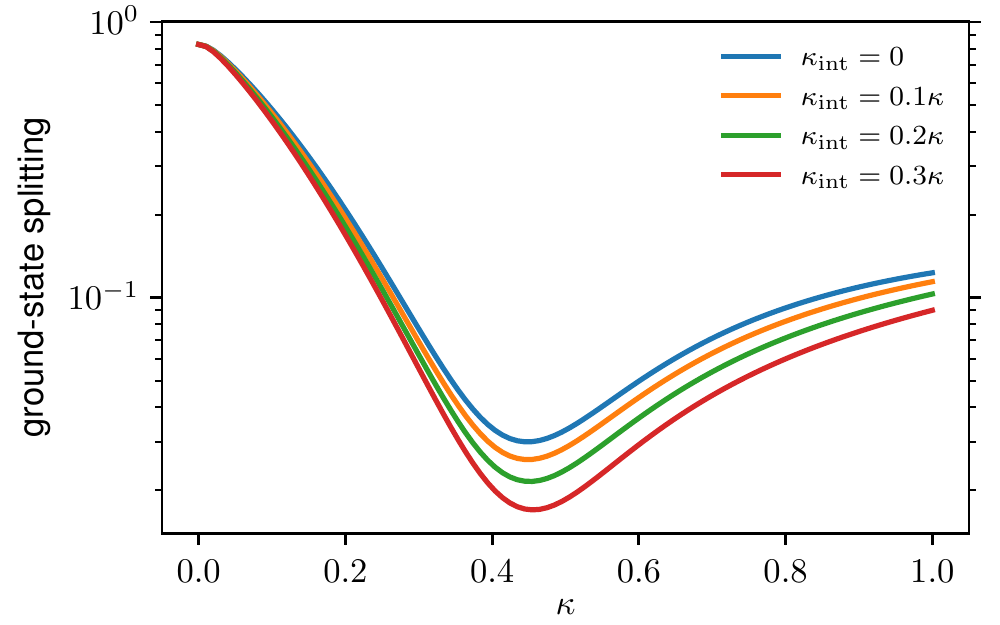}\llap{\raisebox{1.48cm}{\includegraphics[trim = 0 0 -63 0, clip]{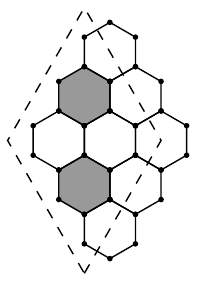}}}
    \caption{The ``ground-state energy splitting" between the ground and the first excited states on a $3\times 3$ honeycomb lattice with two vortices present (shown in gray) as shown in the inset. The energy splitting occurs due to the hybridization of two Majorana modes attached to the vortex excitations. Only in the limit where the vortices are infinitely separated we expect a truly doubly degenerate ground state manifold. This splitting in energy can spoil the braiding properties of the vortices. The size of the splitting depends on a correlation length. The correlation length is bounded from below when we only consider $\kappa$ terms, and the ground-state splitting therefore experiences a minimum around $\kappa \approx 0.4$. The minimal value of the ground-state splitting is further reduced when adding the $\kappa_{\text{int}}$ terms, making the Majorana modes more robust.}
    \label{fig:cor_len_w_majorana_gap}
\end{figure}

When two Majorana modes are close to each other, they can hybridize, which results in a splitting of the ground-state degeneracy. A proxy for the robustness of the two Majorana modes is therefore the size of this energy splitting. 
In Fig.~\ref{fig:cor_len_w_majorana_gap} we calculate the energy splitting as a function of $\kappa$ for a $3\times 3$ honeycomb lattice with periodic boundary conditions in the presence of two vortices. 
Indeed we notice that the gap follows a similar trend to that expected for the correlation length $\xi$ and is bounded from below. 
However, we find that the splitting can be further lowered by adding the $\kappa_{\text{int}}$ terms.
This can be crucial when the calculation is limited in the number of qubits that can be used, but we still want to make the Majorana modes more robust.

\subsection{Dynamical-gauge VQE}\label{sec:mg_part}
\subsubsection{Gauge initial state}
In this section, we consider the case of nonzero, uniform external magnetic fields $\bm h$, as described by Eq.~\eqref{eq:spin_ex_mag}.
In this case, one can no longer restrict the calculation to only one of the $\tilde{\mc L_{u}}$ subspaces, where the configuration of fluxes $W_p$ [see Eq.~\eqref{eq:Wp}] is fixed. 
This follows from Eq.~(\ref{eq:fermion_magnetic_field_hamiltonian}), where the $ib^\alpha_i c_i$ terms flip the sign of the $u_{ij}$ bond variable with $i$ and$j$ making an $\alpha$-edge. 
We thus need to consider the full fermionic Hilbert space $\tilde{\mc L} = \bigoplus_u \tilde{\mc L_{u}}$, which includes all flux sectors. 
To map the system onto qubits, we note that each link variable $u_{ij}$ can be represented by a single qubit. 
A system of $N$ spins thus requires $2N$ qubits to simulate both the flux degrees of freedom and the fermionic (matter) subspace. This qubit overhead limits the system sizes that we can simulate, and we show results up to 8 spins (requiring 16 qubits) below. This suggests using an ansatz formulated in the spin description in the case where the fluxes become dynamic. On the positive side, the simulations in the fermionic description give direct access to nontrivial static properties of the vortex excitations such as their average number in the ground state. The fermionic language is also more natural to use when one is interested in the properties of the Majorana edge modes and their braiding. 

In the last section we discussed how to group the $c_i$ (matter) Majorana fermions into complex fermions, see Eq.~\ref{eq:c_a_mapping}. 
Similarly, the $b_i^\alpha$ (bond) Majorana fermions can be combined into complex fermions in the following way, 
\begin{align}
    g_{(i,j)} = \frac{1}{2} (b^{\alpha}_i + i b^{\alpha}_j), && g^\dagger_{(i,j)} = \frac{1}{2} (b^{\alpha}_i - i b^{\alpha}_j),
\end{align}
where $\alpha = x,y,z$ depending whether $(i,j) \in x,y,z$-edges. Using this basis we can write the gauge variables as 
\begin{align}
    \hat u_{ij} = ib_{i}^{\alpha} b_{j}^{\alpha} = 2 g^\dagger_{(i,j)} g_{(i,j)} - 1, 
\end{align}
and thus initializing a state in a specific gauge configuration amounts to choosing whether a certain fermionic orbital is occupied or empty.

In the same way as we label the sites of the model with Latin indices $i, j, k, \dots$, we will label the edges using Greek letters $\mu, \nu, \lambda, \dots$. However, there is an ambiguity when writing $g_{\mu}$ for example since $g_{(i,j)} \neq g_{(j,i)}$, but $(i,j)$ and $(j,i)$ are the same edge. In order to remove this ambiguity we define $g_{\nu}$ such that $g^\dagger_\nu g_\nu = 1$ on all edges corresponding to the standard configuration $u^{\text{std}}_{ij}$, shown in Fig.~\ref{fig:def_lattices}(c).

\begin{figure*}[t]
    \centering
    \subfloat[$h = h_0(1,1,1)$]{\includegraphics[trim = 2 0 7 0, clip, scale=0.6]{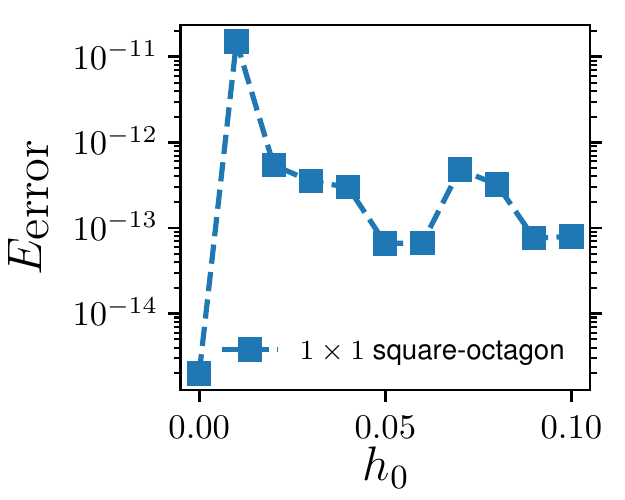} }
    \subfloat[$h = h_0(1,1,1)$]{\includegraphics[trim = 8 0 7 0, clip, scale=0.6]{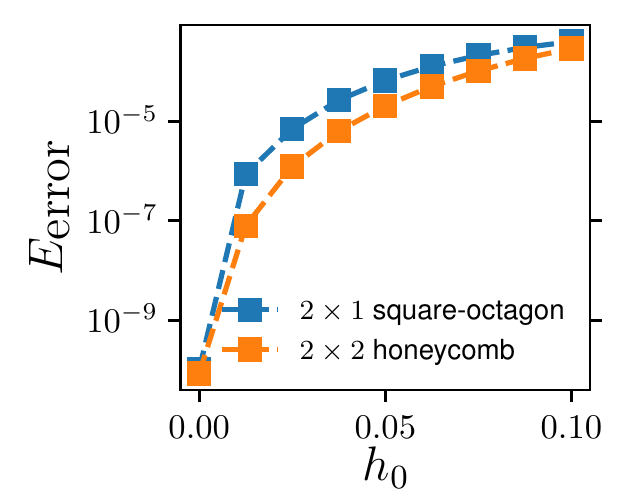} }
    \subfloat[Average polarization. $h = (0,0,h_z)$.]{\includegraphics[trim = 2 0 7 0, clip,scale=0.6]{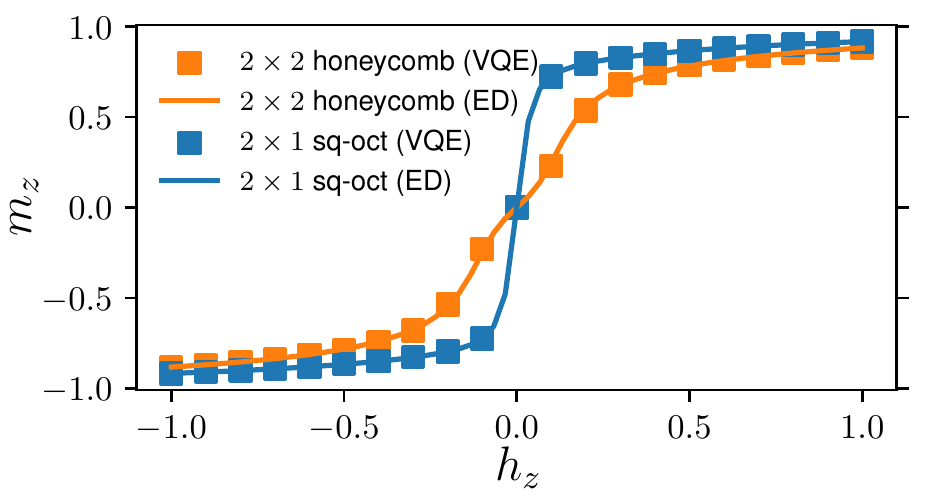}}
    \subfloat[Average plaquettes. $h = (h_x,0,0)$.]{\includegraphics[trim = 20 0 3 0, clip,scale=0.6]{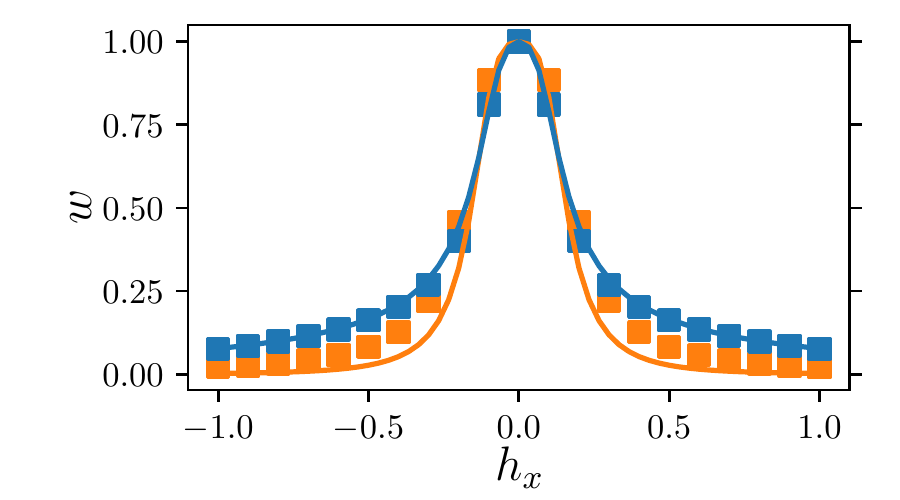}}
    \caption{Dynamic gauge VQE results using a statevector simulator. Panels (a) and (b) show the VQE energy error compared to exact diagonalization (ED), $E_{\text{error}} =  \abs{(E_{\text{VQE}} - E_{\text{exact}})/E_{\text{exact}}   }$, as a function of magnetic field. The accuracy is close to machine precision for the $1 \times 1$ square octacon model, and increases as a function of $h_0$ up to $\approx 10^{-3}$ for the larger models. Panel (c) shows the ground state magnetization $m_z = \sum_i \expval{\sigma^z_i}/N$ as a function of field applied along the $z$ axis for both honeycomb and square-octagon models. It shows very good agreement with exact results, and shows that an external field induces a finite magnetization $m_z$ that saturates when the field becomes of the order of the Kitaev exchange. The susceptibility is larger for the square-octagon model. Panel (d) shows the average value of the plaquette operator $w = \sum_p \expval{W_p}/n$ ($n$ is the total number of plaquettes) in the ground state as a function of a magnetic field $h_x$ applied along $x$. The legend is identical to panel (c).  
    Results show that fluxes proliferate due to an external field, and they give a quantitative estimate to the extent of the perturbative regime, where one considers only the flux free sector with $w = 1$. For all calculations we use $\bm J = (1,1,1)$ for the honeycomb lattice and $\bm J = (1,1,\sqrt{2})$ for the square-octagon lattice.}
    \label{fig:dynamic_gauge_vqe_results}
\end{figure*}

We choose to initialize the system in the standard gauge configuration, with all $u_{ij} = 1$.
{Thus our} initial state is such that
\begin{align}
    g^\dagger_{\nu} \ket{\psi_0} = 0, \quad \forall g^\dagger_{\nu} 
\end{align}

\subsubsection{Variational ansatz}
We use the following ansatz when extending the calculation to include all gauge configurations, 
\begin{align}
    \ket{\psi(\bm \theta)} =  \exp\[\sum \theta^c_{ij} c_ib_j\] \exp\[ \sum \theta^b_{ij} b_ib_j\] \nonumber \\ \qquad  \qquad  \qquad \qquad   \times \exp\[ \sum \theta^a_{ij} c_ic_j\] \ket{\psi_0} \nonumber \\ 
     \equiv U^c(\bm \theta^c) U^b(\bm \theta^b) U^a(\bm \theta^a) \ket{\psi_0}. 
    \label{eq:ansatz_full}
\end{align}
Similar to the discussion in Sec.~\ref{sec:fg_ansatz}, and Appendix~\ref{app:relevant_ansatz}, we choose to reduce the number of parameters by keeping only the following terms, 
\begin{align} \label{eq:dgvqe_u_a}
    &U^a(\bm \theta^a) = \nonumber \\ 
    &\prod_{ij} \exp \[ i \theta^{a_1}_{ij}(a^\dagger_i a^\dagger_j + a_ja_i)\] \exp\[ \theta^{a_2}_{ij}(a^\dagger_i a^\dagger_j - a_ja_i) \],  \\  \label{eq:dgvqe_u_b}
    &U^b(\bm \theta^b) = \nonumber \\ 
    &\prod_{\mu \nu} \exp \[ i \theta^{b_1}_{\mu \nu}(g^\dagger_{\mu} g^\dagger_{\nu} + g_{\nu}g_\mu) \] \exp \[ \theta^{b_2}_{\mu \nu}(g^\dagger_\mu g^\dagger_\nu - g_\nu g_\mu) \],  \\ \label{eq:dgvqe_u_c}
    &U^c(\bm \theta^c) = \nonumber \\ 
    &\prod_{i\mu} \exp \[ i \theta^{c_1}_{i\mu}(a^\dagger_i g^\dagger_\mu + g_\mu a_i) \] \exp \[ \theta^{c_2}_{i\mu}(a^\dagger_i g^\dagger_\mu - g_\mu a_i) \] .
\end{align}
Here, $\bm \theta^a$ contains $\frac{N}{2}\bigl(\frac{N}{2} -1\bigr)$ parameters for a system with $N$ spins. Such a system will have $\frac{3N}{2}$ edges, and thus $\bm \theta^b$ contains $\frac{3N}{2} \bigl(\frac{3N}{2} - 1\bigr)$ parameters, and $\bm \theta^c$ contains $\frac{3N^2}{2}$ parameters. In total, the ansatz $U(\bm \theta)$ has $2N (2N - 1)$ parameters.

\subsubsection{Avoiding unphysical states}
\label{sec:cost_fun}
The variational state in Eq.~(\ref{eq:ansatz_full}}) explores states in the full Hilbert space, which includes both physical and unphysical states. When using the expectation value of the energy $C(\bm \theta) = \expval*{H}{\psi(\bm \theta)}$ as a cost function, it is not guaranteed that the optimal state $\ket{\psi(\bm \theta_{\text{optimal}})}$ belongs to the physical subspace.  
Unphysical states are defined such that $\mc P \ket{\psi} = 0$, where $\mc P$ is the projection operator as defined in Eq.~(\ref{eq:full_projection_op}).
Therefore, in the case of VQE with a dynamical gauge field we use the following cost function
\begin{align}\label{eq:mod_cost}
    C(\bm \theta) = \frac{\expval*{\mc P H}{\psi(\bm \theta)} }{\expval*{\mc P}{\psi(\bm \theta)}},
\end{align}
which explicitly includes the projector onto the physical subspace. We observed that using this cost function, the algorithm always converged to a physical state in all the cases tested. 

In Appendix~\ref{app:JW} we discuss the Jordan-Wigner transformation of the Majorana fermions of the Kitaev model. The Jordan-Wigner transformation of the projection operator $\mc P$ (much like the Hamiltonian) is a sum of Pauli strings. 
Since both $\mc P$ and $H$ are a sum of Pauli strings, $\mc P H$ is also a sum of Pauli strings. 
Pauli strings are observables that can be measured on a quantum computer, and how to measure sum of Pauli strings efficiently has been discussed in the literature~\cite{pauli_mes_2021, yen2020measuring, jena2019pauli, Izmaylov_2020, love_2020}. 

Another possible solution to making sure $\ket{\psi(\bm \theta_{\rm optimal})}$ belongs to the physical subspace is to modify the cost function by adding an on-site chemical potential to the Hamiltonian $C(\bm \theta) = \expval{ H -\mu \sum_i \mc P_i}$. 
Such a chemical potential term penalizes states for having an unphysical component. 
On one hand this solution has the advantage of only needing to measure local projectors $\mc P_i$ instead of the full projection $\mc P = \prod_i \mc P_i$ which can get very small for large system sizes and can limit the scalability of the method above. 
On the other hand because $\[ H, \mc P_i\]\neq 0$ the addition of this chemical potential term will make the optimization process harder, and probably one will need to fine tune $\mu$ for optimal results. 
It is an interesting problem to compare both methods in more detail, and we leave it for future work. In this work, for the sake of demonstrating our ansatz, we only use Eq~(\ref{eq:mod_cost}). 

\subsubsection{Simulations and results}

With the gauge variables being dynamic, we demonstrate the capabilities of the ansatz above using three geometries: $1 \times 1$ square-octagon, $2 \times 1$ square-octagon, and $2 \times 2$ honeycomb lattice.
These geometries have $4$, $8$ and $8$ spins, respectively, and thus require $8$, $16$, and $16$ qubits to simulate. 
Periodic boundary conditions are applied in both cases.
As before, we set both models inside the gapless region of the phase diagram when the magnetic field vanishes.
For the honeycomb lattice we set $J = (1,1,1)$, and for the square-octagon lattice we set $J = (1,1,\sqrt{2})$. 

Fig.~\ref{fig:dynamic_gauge_vqe_results} (a) and (b) show the error in the ground state energy comparing the VQE results to that of exact diagonalization in the presence of a uniform magnetic field $\bm h = h_0(1,1,1)$. 
Again we compare our results to those in Ref.~\cite{Andy_2021}. 
Here we can make more direct comparisons since we are simulating the same added interactions to the Kitaev model $i.e.$ external fields. 
We expect that for small enough value of $h_0$ our method should always perform better. For $h_0 = 0.05/\sqrt{3}$, the value of the field studied in Ref.~\cite{Andy_2021}, we find similar $E_{\text{error}}$ between our method and the best method described in Ref.~\cite{Andy_2021} of about $10^{-6}$.

Unlike the fixed gauge VQE case, in this case it is hard to compare the optimized state infidelity with respect to exact diagonalization. The reason for this is the massive degeneracy introduced by the gauge freedom.
Adding the magnetic field terms does not change the fact that the model is invariant under a gauge transformation. 
Thus the full fermionic Hilbert space has many degenerate ground states that can be related to each other by a gauge transformation. 
This makes comparing state fidelity much harder than in the fixed-gauge VQE case, especially for the $16$-qubit cases where getting the full spectrum using exact diagonalization is time consuming and we could only solve for the ground state even in the exact diagonalization calculation. 
There is no guarantee that the ground state found by exact diagonalization should be the same as the ground state found by the VQE algorithm. 
We do expect them to be gauge related, though. 

We validate the accuracy of the optimized state by calculating some known physical features that is gauge independent. In Fig.~\ref{fig:dynamic_gauge_vqe_results}(c) we show the average polarization $m_z = \sum_i \expval{\sigma^z_i}/N$, and in Fig.~\ref{fig:dynamic_gauge_vqe_results}(d) we present the average value of the plaquette operator $w = \sum_p \expval{W_p}/n$. Here, $n$ is the total number of plaquettes. 
Both quantities show good agreement between our VQE results (square markers) and exact diagonalization (solid lines). We note that similar calculation where shown in Ref.~\cite{Kyriienko_2021}.

For the average magnetization shown Fig.~\ref{fig:dynamic_gauge_vqe_results} (c) we see that the magnetization vanishes at zero field, $m_z(\bm h = 0) = 0$, which is a signature of the spin-liquid phase. As the magnitude of the field increases the magnetization increases until it reaches a saturation point where all spins are pointing in the same direction as the field (indicating a fully polarized paramagnetic state). At least for the geometries considered, we see that the magnetic susceptibility $\eval{\dfrac{\partial m_z}{\partial h_z}}_{h_z = 0}$ is larger for the square-octagon than for the honeycomb lattice model. 

Fig.~\ref{fig:dynamic_gauge_vqe_results} (d) shows that at $\bm h = 0$ the ground state has $w = 1$ as expected since all $W_p = 1$ (no vortices). As more and more flux is put through the system $w$ decreases as vortices are excited in the system. At large values of the field in the $x$-direction we see $w$ going to zero. This is consistent for both the honeycomb and square-octagon lattices where a product state with all spins pointing in the $x$-direction would yield $W^{(6)}_p  = W^{(4)}_p  = W^{(8)}_p  = 0$. This is different in the situation where the field is pointing along the $z$-direction since a state that is polarized along $z$ would still have $W^{(4)}_p = 1$ for the square-octagon lattice.   


\section{Conclusion}
We simulate Kitaev spin models using VQE with an ansatz that is motivated by the fermionic description of the model. 
In cases where the gauge degrees of freedom are static, our method only requires half as many qubits as there are physical spins in the model. This includes nontrivial cases where the matter fermion problem is interacting. Such interaction terms arise when treating an external magnetic field within perturbation theory. We show that using the fermionic formulation has the additional advantage of being able to realize and simulate properties of non-Abelian anyons, i.e., Majorana zero modes bound near static vortex excitations. Specifically, we find that the Majorana bound states can become more localized (and thus more robust) in the presence of matter fermion interactions.
We can capture these excited states within a ground state VQE calculation by running a matter fermion VQE on top of a static background of vortex excitations. We see this as an exciting new direction that can be explored in more detail in future work. 

We find that the accuracy of our method generally compares well to other VQE studies of the Kitaev model on the square-octagon lattice. In the presence of a uniform external magnetic field sufficiently small that it can be treated within perturbation theory, our method shows at least one order of magnitude better $E_{\text{error}}$ for the 16 spin geometry than results presented in Ref.~\cite{Andy_2021}. 
The better $E_{\text{error}}$ in this work demonstrates that fermionizing spin models can provide an advantage when additional constraints limit the size of the Hilbert space where the ground state is located. 

Further, we expand our method to perform VQE simulations in the presence of terms that couple different gauge sectors. 
In this case we encountered a challenging issue that VQE converged to completely unphysical states. 
We offered two possible solutions to this problem: we used one of them in this work, and leave the other one for future work. 
Optimization over a set of constraints is an interesting problem in its own right, and having a separate future study comparing the various ways of handling the problem for our method is useful.
Future work could also be directed towards performing an in-depth comparison between VQE ans\"atze in the fermionic and the spin description (such as the Hamiltonian variational ansatz used in Refs.~\cite{Andy_2021,Kyriienko_2021}) with regards to the depth of the circuits and the complexity of the classical optimization, in particular in the presence of noise.

\begin{acknowledgments}
This material is based upon work supported by the U.S. Department of Energy, Office of Science, National Quantum Information Science Research Centers, Superconducting Quantum Materials and Systems Center (SQMS) under the contract No. DE-AC02-07CH11359.
We would like to thank the entire SQMS algorithms team for fruitful and thought provoking discussions around this work.
In particular we would like to thank A. B. \"Ozg\"uler and S. Hadfield for a thoughtful review of this manuscript.
\end{acknowledgments}
\appendix

\section{Relevant parts of the ansatz}
\label{app:relevant_ansatz}
Our ansatz introduced in Sec.~\ref{sec:fg_ansatz} can be simplified by modding out the parts of the ansatz that leaves the initial state invariant. Let us look at the action of $U^a(\bm \theta^a)$ on $\ket{\psi_0}$, where
\begin{align}\label{eq:Ua}
    U^a(\bm \theta^a) = \exp\[ \sum_{ij} \theta^a_{ij} c_i c_j \],
\end{align}
and $a_i \ket{\psi_0} = 0$ for all $a_i$.  
We begin by writing $U^a(\bm \theta^a)$ in terms of the complex fermions $a_i$,
\begin{align}
    &U^a(\bm \theta^a) = \exp\sum_{ij} (\theta^a_{2i, 2j} c_{2i}c_{2j} + \theta^a_{2i+1, 2j} c_{2i+1}c_{2j}  \nonumber \\
    & \qquad \quad + \theta^a_{2i, 2j+1} c_{2i}c_{2j+1} +\theta^a_{2i+1, 2i+1} c_{2i+1}c_{2j+1}) \nonumber \\
    &= \exp\[\sum_{ij} \[ i \theta^{a_1}_{ij}(a^\dagger_i a_j + a^\dagger_j a_i) +  \theta^{a_2}_{ij}(a^\dagger_i a_j - a^\dagger_j a_i) \]\right.  \nonumber \\ 
    &\  \quad  \left. + \sum_{ij} \[ i \theta^{a_3}_{ij}(a^\dagger_i a^\dagger_j + a_ja_i) + \theta^{a_4}_{ij}(a^\dagger_i a^\dagger_j - a_ja_i) \] \], 
\end{align}
where it can be shown that, 
\begin{align}
    \theta^{a_1}_{ij} = \theta^a_{2i,2j} - \theta^a_{2i+1, 2j+1}  \nonumber \\ 
    \theta^{a_2}_{ij} = \theta^a_{2i,2j} + \theta^a_{2i+1, 2j+1} \nonumber \\ 
    \theta^{a_3}_{ij} = \theta^a_{2i+1,2j} + \theta^a_{2i, 2j+1}  \nonumber \\ 
    \theta^{a_4}_{ij} = \theta^a_{2i + 1,2j} - \theta^a_{2i, 2j+1}
\end{align}
Since the commutators $[a^\dagger_{i} a_{j},a^\dagger_{i'} a_{j'}]$, $[a^\dagger_{i} a_{j},a_{i'} a_{j'}]$, $[a^\dagger_{i} a^\dagger_{j},a_{i'} a_{j'}]$, and $[a_{i} a_{j},a_{i'} a_{j'}]$ are either zero or a quadratic product of $a_i$'s and $a^\dagger_i$'s we can write 
\begin{align}
    & \exp\[\sum_{ij} \[ i \theta^{a_3}_{ij}(a^\dagger_i a_j + a^\dagger_j a_i) +  \theta^{a_4}_{ij}(a^\dagger_i a_j - a^\dagger_j a_i) \]\right.  \nonumber \\ 
    & \quad  \left. + \sum_{ij} \[ i \theta^{a_1}_{ij}(a^\dagger_i a^\dagger_j + a_ja_i) + \theta^{a_2}_{ij}(a^\dagger_i a^\dagger_j - a_ja_i) \] \] \nonumber \\ 
    =&\prod_{ij} \exp \[ i \theta^{'a_1}_{ij}(a^\dagger_i a^\dagger_j  + a_ja_i) \]  \exp \[ \theta^{'a_2}_{ij}(a^\dagger_i a^\dagger_j - a_ja_i) \]   \nonumber \\ 
    \times & \prod_{ij} \exp \[ i \theta^{'a_3}_{ij}(a^\dagger_i a_j + a^\dagger_j a_i) \] \exp \[  \theta^{'a_4}_{ij}(a^\dagger_i a_j - a^\dagger_j a_i) \]
\end{align}
However since the parameters in the exponent are to be found variationally anyway, the exact relationship between the primed and unprimed $\theta$'s is not relevant, and we can just as well use the RHS of the equation above in our ansatz. Finally we notice that, 
\begin{align}
    \prod_{ij} \exp &\[ i \theta^{a_3}_{ij}(a^\dagger_i a_j + a^\dagger_j a_i) \] \exp \[  \theta^{a_4}_{ij}(a^\dagger_i a_j - a^\dagger_j a_i) \] \ket{\psi_0} \nonumber \\ 
    &= e^{i\phi} \ket{\psi_0}
\end{align}
since $a_i \ket{\psi_0} = 0$ for all $a_i$. Thus in our ansatz we only use 
\begin{align}
    &U^a(\bm \theta^a) \equiv \nonumber \\ 
    & \prod_{ij} \exp \[ i \theta^{a_1}_{ij}(a^\dagger_i a^\dagger_j  + a_ja_i) \] \exp \[ \theta^{a_2}_{ij}(a^\dagger_i a^\dagger_j - a_ja_i) \] 
\end{align}
without any loss of generality. For a system with $N$ spins, the above expression has $N/2(N/2 - 1)$ as opposed to the $N(N-1)/2$ independent parameters of Eq.~(\ref{eq:Ua}). 

\section{Mapping the fermionic model onto qubits}\label{app:JW}
\subsection{The Jordan-Wigner transformation}
A system of $N$ qubits has a $2^N$ dimensional Hilbert space that is spanned by, 
\begin{align}
    \ket{s_1 \dots s_N} = (\sigma_1^+)^{s_1} \dots (\sigma_N^+)^{s_N} \ket{0}, && s_i \in \{0,1\},
\end{align}
where $\sigma^+_i = \frac{1}{2} (\sigma^x_i - i\sigma^y_i)$, and the state $\ket{0}$ is defined such that $\sigma^-_i \ket{0} = 0$ for all $\sigma_i^- = (\sigma^+_i)^\dagger$. The operators $\sigma^\pm$ obey the following commutation relationships
\begin{align}\label{eq:boson_algebra}
    \[\sigma^-_i, \sigma^+_j \] = \delta_{ij}, && \[\sigma^-_i, \sigma^+_j\] = 0. 
\end{align}

Consider a fermionic Hilbert space of $N$ orbitals (we take orbital here to also include the spin) with $a_i, \ i \in \{ 1,\dots N\}$ being the annihilation operators for these $N$ orbitals. Like qubits the Hilbert space is $2^N$ dimensional, since each orbital can either be full or empty, and is spanned by 
\begin{align}
    \ket{n_1 \dots n_N} = (a_1^\dagger)^{n_1} \dots (a_N^\dagger)^{n_N} \ket{0},&& n_i \in \{0,1\},
\end{align}
However unlike qubits the operators $a_i$'s satisfy the following anti-commutation relationships, 
\begin{align}\label{eq:fermi_algerba}
    \{a_i, a^\dagger_j\} = \delta_{ij}, && \{a_i, a_j\} = 0
\end{align} 
where $\ket{0}$ is the state annihilated by all lowering operators.


The difference in the algebra described by Eqs.~(\ref{eq:boson_algebra}) and~(\ref{eq:fermi_algerba}) prevents a simple map such as  $\sigma_i^- = a_i$, and  $\sigma_i^+ = a_i^\dagger$. 
Instead, the Jordan-Wigner transformation yields the correct mapping that preserves the correct anti-commutation relationships of the fermions, 
\begin{align}\label{eq:complex_fer_JW}
    a_i = \prod_{j<i} \sigma^z_j \sigma^-_i, && a^\dagger_i = \prod_{j<i} \sigma^z_j \sigma^+_i.  
\end{align}

\subsection{Transforming the Hamiltonian and the ansatz}
\begin{figure*}
    \centering
    \includegraphics[scale = 0.81]{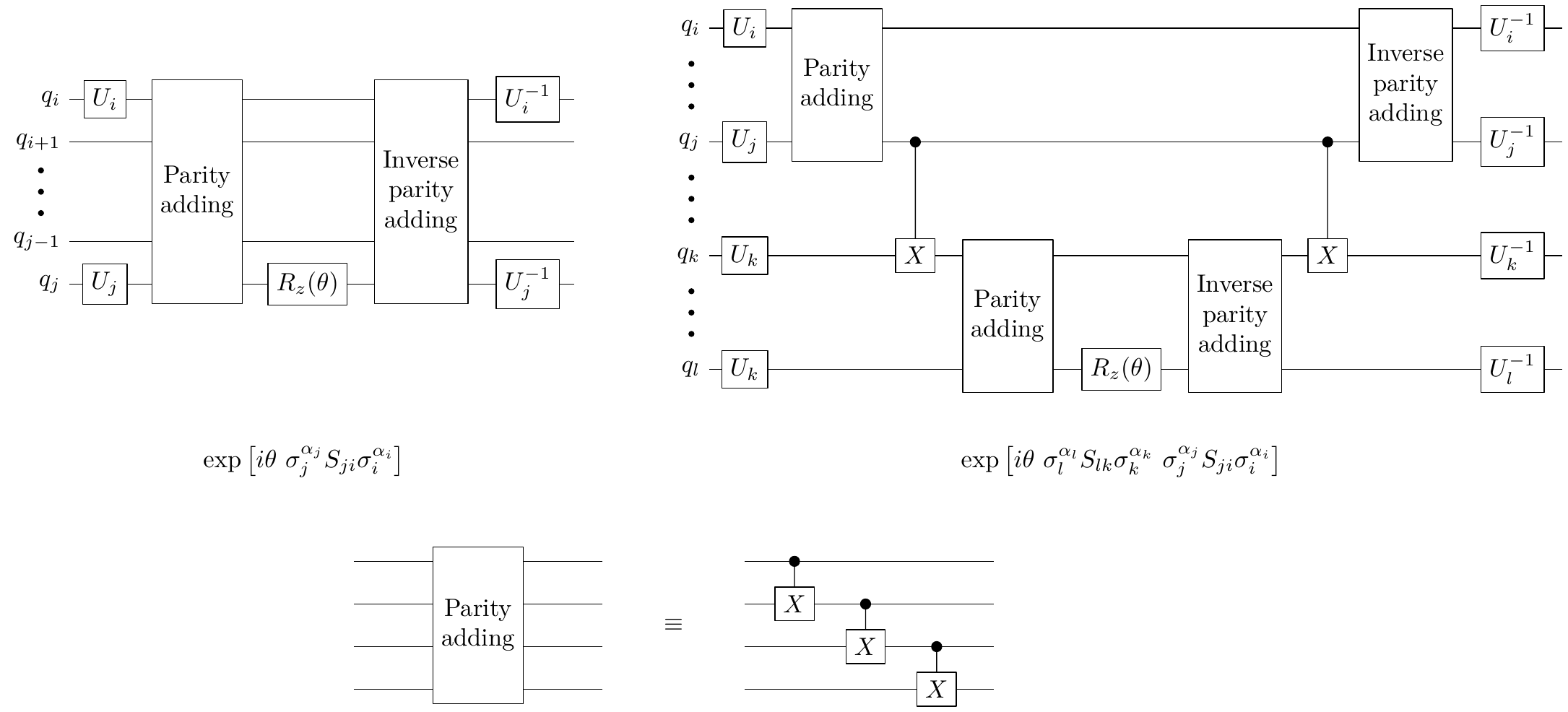}
    \caption{Building blocks for the ansatz. Here $U_i= H$ (Hadamard gate) if $\alpha_i = x$, and $U_i = R_x(\frac{\pi}{2})$ (rotation about $x$-axis by $\pi/2$) if $\alpha_i = y$. }
    \label{fig:ansatz_form}
\end{figure*}
As discussed in the main text, it is useful to have a distinction between two kinds of Majoranas of the Kitaev model, the $b^\alpha_i$ Majoranas that make up the gauge sector and the $c$ Majoranas that make up the fermionic sector. 
In the main text we chose to have, 
\begin{align}
    c_{2n} = a_n + a_n^\dagger, && c_{2n + 1} =  \frac{1}{i}(a_n - a^\dagger_n).
\end{align}
Since a pair of Majoranas combine to make a complex fermion, for a system of $N$ spins the index $n$ above ranges from $1$ to $N/2$. Using the transformation in Eq.~(\ref{eq:complex_fer_JW}) we see that the Majorana fermions maps to the following, 
\begin{align}\label{eq:c_JW}
    c_{2n} = \prod_{m<n} \sigma^z_m \sigma^x_n && c_{2n + 1} = \prod_{m<n} \sigma^z_m \sigma^y_n.
\end{align}
Further, we also have a complex fermion $g_\mu$ associated with each edge as discussed in the main text. Since the complex fermions $g_\mu$ are defined in such a specific way such that $g^\dagger_\mu g_\mu = 1$ corresponds to the standard gauge configuration $u^{std}_{ij}$, we need a new notation for the $b^\alpha_i$ Majorana fermions in order to avoid ambiguous notations and properly keep track of minus signs. We define,
\begin{align}
    b_\nu^{1} = g^\dagger_{\nu} + g_{\nu} && b_\nu^{2}  = \frac{1}{i} (g^\dagger_{\nu} - g_{\nu}),
\end{align}
For the Jordan-Wigner transformation we make the following identification, 
\begin{align}
    g_\nu \equiv a_{\nu + N/2}.
\end{align}
With this we can extend the Jordan-Wigner transformation to include the $b^\alpha_i$ Majorana Fermions
\begin{align}\label{eq:b_JW}
    b^1_{\nu} = \prod_{m<\nu + N/2} \sigma^z_m \sigma^x_{\nu + N/2} && b^2_\nu = \prod_{m<\nu + N/2} \sigma^z_m \sigma^y_{\nu + N/2}.
\end{align}

Using Eqs.~(\ref{eq:c_JW}) and~(\ref{eq:b_JW}) one can work out the Jordan-Wigner transformation of all possible terms in the Hamiltonian. 
Defining 
\begin{align}
    S_{ji} = \prod_{j  \le  p< i}  \sigma^z_{p}
\end{align}
the fixed gauge Hamiltonian transforms as follows,
    \begin{align}
        & \ \ \sum_{j>i} i A_{ij} c_i c_j = \sum_{j>i} A_{ij} \  i \sigma^{\alpha_i}_{i'}   S_{i'  j'}   \sigma^{\alpha_j}_{j'},  \\ 
        &\sum_{l>k>j>i} V_{ijkl} c_i c_j c_k c_l  = 
        \sum_{l>k>j>i} \sigma^{\alpha_i}_{i'}  S_{i'j'}  \sigma^{\alpha_j}_{j'}  \  \sigma^{\alpha_k}_{k'}  S_{k'l'}   \sigma^{\alpha_l}_{l'},
    \end{align}
with $i',j',k',l' = \lfloor i / 2 \rfloor, \lfloor j / 2 \rfloor, \lfloor k / 2 \rfloor,  \lfloor l / 2 \rfloor $,  $\alpha_i = x$ when $i$ is even, and $\alpha_i = y$ when $i$ is odd. 

When dealing with dynamic gauge Hamiltonian we have, 
    \begin{align}
        \sum_{j>i} J_\alpha c_i c_j b^\alpha_i b^\alpha_j = i \sigma^{\alpha_i}_{i'} S_{i'j'} \sigma^{\alpha_j}_{j'} \  \[ s_{ij}  \sigma^z_{\nu + N/2} \], \\
        \sum_{i} h_\alpha   c_i b_i^\alpha = i \sigma^{\alpha_i}_{i'} S_{i' \nu + N/2}  \  \sigma^{\beta_{\alpha}}_{\nu + N/2}
    \end{align}
where $s_{ij} = \pm$ , and $\beta_{\alpha} = x, y$ when $b^\alpha_{i} = b^1_\nu$, or $b^\alpha_{i} = b^2_\nu$ respectively.

We now move on to the Jordan-Wigner transformed ansatz. We start with
\begin{align}
    &U^a(\bm \theta^a) \equiv \nonumber \\ 
    & \prod_{i<j} \exp \[ i \theta^{a_1}_{ij}(a^\dagger_i a^\dagger_j  + a_ja_i) \] \exp \[ \theta^{a_2}_{ij}(a^\dagger_i a^\dagger_j - a_ja_i) \]. 
\end{align}
Using Eq.~\ref{eq:complex_fer_JW}, we can write the exponents as
\begin{align} \label{eq:U_a_JW}
    & a^\dagger_i a^\dagger_j  + a_ja_i =  2(\sigma^x_i S_{ij} \sigma^x_j  - \sigma^y_i S_{ij} \sigma^y_j ) \\ \label{eq:U_b_JW}
    & a^\dagger_i a^\dagger_j  - a_ja_i =
     -2i( \sigma^x_i S_{ij} \sigma^y_j  + \sigma^y_i S_{ij}  \sigma^x_j  )
\end{align}
Next we look into the stransformation of 
\begin{align}
    &U^b(\bm \theta^b) \equiv \prod_{i<j<k<l} \exp \[  i \theta^{b_1}_{ijkl}(a^\dagger_i a^\dagger_ja^\dagger_k a^\dagger_l   + a_la_ka_ja_i) \] \nonumber \\
    &  \qquad \qquad    \times \exp \[ \theta^{b_2}_{ijkl}(a^\dagger_i a^\dagger_ja^\dagger_k a^\dagger_l - a_la_ka_ja_i) \] .
\end{align}
Using Eq.~\ref{eq:complex_fer_JW} the exponents can be transformed as follows,
\begin{align}
    a^\dagger_n a^\dagger_ma^\dagger_k a^\dagger_l   + a_la_ka_ma_n = 
    \(\sigma^x_i S_{ij} \sigma^x_j \ \sigma^x_k S_{kl} \sigma^x_l \right. \nonumber \\
    - \sigma^y_i S_{ij} \sigma^y_j  \ \sigma^x_k S_{kl} \sigma^x_l -  \text{all permutations of } x, y   \nonumber \\
    \left.  +  \sigma^y_i S_{ij} \sigma^y_j  \ \sigma^y_k S_{kl} \sigma^y_l  \)  \\ 
    a^\dagger_n a^\dagger_ma^\dagger_k a^\dagger_l   - a_la_ka_ma_n = 2i \( \qquad  \qquad  \qquad  \ \  \right.   \nonumber \\ 
    \sigma^y_i S_{ij} \sigma^y_j \ \sigma^y_k S_{kl} \sigma^x_l + \text{all permutations of } x, y \nonumber \\
    - \left. \sigma^x_i S_{ij} \sigma^x_j \ \sigma^x_k S_{kl} \sigma^y_l + \text{all permutations of } x, y  \). 
\end{align}
In Fig.~\ref{fig:ansatz_form} we show how this transformed ansatz can be implemented on a quantum computer. 
Finally, we note that the ansatz used in the dynamical gauge VQE (Eqs.~(\ref{eq:dgvqe_u_a}, \ref{eq:dgvqe_u_b}, \ref{eq:dgvqe_u_c})) can be transformed to operators that can be acted with on qubits using equations that are very similar to Eqs.~(\ref{eq:U_a_JW}) and (\ref{eq:U_b_JW}).

\bibliographystyle{apsrev4-2}
\bibliography{main}
\end{document}